\documentclass[sigconf, nonacm]{acmart}
\usepackage{subcaption}
\usepackage{enumitem}
\usepackage[para]{threeparttable}
\usepackage{multirow}
\usepackage[linesnumbered,ruled,vlined]{algorithm2e}
\usepackage{booktabs}

\graphicspath{{figures/}}

\begin{document}
\title{Eiger: An Efficient Library for GPU-based Data Analytics}

%%
%% The "author" command and its associated commands are used to define the authors and their affiliations.
\author{Bowen Wu}
\affiliation{%
  \institution{Systems Group, ETH Zurich}
  \country{Switzerland}
}
\email{bowen.wu@inf.ethz.ch}

\author{Marko Kabić}
\affiliation{%
    \institution{Systems Group, ETH Zurich}
  \country{Switzerland}
}
\email{marko.kabic@inf.ethz.ch}

\author{Sven Hepkema}
\affiliation{%
    \institution{Systems Group, ETH Zurich}
  \country{Switzerland}
}
\email{sven.hepkema@inf.ethz.ch}

\author{Vasilis Mageirakos}
\affiliation{%
    \institution{Systems Group, ETH Zurich}
  \country{Switzerland}
}
\email{vmageirakos@inf.ethz.ch}

\author{Christos Kozyrakis}
\affiliation{%
    \institution{NVIDIA \& Stanford University}
  \country{United States}
}
\email{ckozyrakis@nvidia.com}

\author{Gustavo Alonso}
\affiliation{%
    \institution{Systems Group, ETH Zurich}
  \country{Switzerland}
}
\email{alonso@inf.ethz.ch}

\begin{abstract}
GPUs have become an increasingly attractive platform for accelerating analytical workloads due to their massive parallelism and high memory bandwidth. Recent studies show that in systems with fast CPU-GPU interconnects and fast networks, query processing within the GPU, rather than data movement, is the dominant bottleneck. This highlights the need for a more efficient implementation of relational operators on GPUs than the widely used library, cuDF. While offering rich functionality, cuDF commits to a single, statically chosen implementation for most operators and makes little use of runtime information about the data, limiting performance across diverse workloads and different GPUs. In this paper, we present Eiger, a high-performance library for GPU-based data analytics that improves single-GPU query processing through \emph{runtime workload adaptivity}. Adaptivity in Eiger rests on two design principles. First, Eiger provides multiple implementation variants and tunable knobs for most operators, covering not only joins and group-bys but also expensive yet often overlooked operations, such as expression evaluation, string processing, and multi-key sorting, for which it contributes an array of new optimization techniques. Second, Eiger profiles intermediate data \emph{during query execution} using lightweight statistics, such as value ranges and HyperLogLog++ sketches, and uses them to select implementations, tune configuration knobs, and compress data on the fly, thereby overcoming the limitations of traditional static query optimization. The breadth of operators and variants also enables a more comprehensive performance analysis, encompassing a wider range of operations and workloads than previous work.
We evaluate Eiger with operator microbenchmarks on two GPU architectures and with the complete TPC-H benchmark (up to scale factor 100). Across the 22 queries, Eiger reduces the total runtime by up to 1.8$\times$ compared to the state-of-the-art cuDF library; for individual queries, Eiger achieves up to 6.1$\times$ better performance.
\end{abstract}

\maketitle

\section{Introduction}
Recent years have seen rapid growth in the use of graphics processing units (GPUs) to accelerate database applications~\cite{wu2025terabytescaleanalyticsblinkeye,yogatama_rethinking_2025,surakav_he_2022}. GPUs, with their massive parallelism and high-bandwidth memory, are well-suited for many, if not all, operations in data analytics. Recently, driven by large language model training and inference, GPUs have been deployed at large scales and are equipped with increasingly higher memory bandwidth, larger memory capacity, faster CPU-GPU interconnects, and high-bandwidth RDMA networks. All of these enable GPU-based databases to process data at one to two orders of magnitude faster than CPU-based databases~\cite{wu2025terabytescaleanalyticsblinkeye,yogatama_rethinking_2025}. 

Recent performance studies by Kabić et al.~\cite{kabic25-maxbench} and Wu et al.~\cite{wu2025terabytescaleanalyticsblinkeye} evaluate GPU-based database systems on two different deployments: (1) relations initially stored in CPU DRAM requiring data transfer to the GPU, and (2) relations pre-partitioned and loaded into high-bandwidth memory (HBM) of multiple GPUs in a cluster. Across both settings using the TPC-H benchmark, they find that query processing time on the GPU dominates overall execution time when fast interconnects (e.g., NVLink-C2C) and RDMA networks (e.g., InfiniBand) are available. This challenges the common assumption that data movement is the main bottleneck and suggests that the cost of query execution on GPUs will become even more critical as faster connectivity becomes mainstream~\cite{microsoft-nvl72,microsoft-rubin}. These observations motivate research focused on improving \emph{single-GPU query processing performance}.

Existing systems are typically built on cuDF~\cite{cudf}, an open-source GPU DataFrame library, or on tensor operators from machine learning libraries~\cite{surakav_he_2022,zhang25-tqex}. These libraries commit to a single, statically chosen implementation for most operators. cuDF, for instance, executes both joins and group-bys with non-partitioned hash-based implementations by default. This one-size-fits-all design is at odds with decades of CPU database experience, which shows that no single implementation performs best for all workloads~\cite{db-cowbook}: hash-based joins and group-bys perform well when the hash table is cache-resident, but degrade otherwise, where sort- or partition-based methods are faster. We identify the lack of \emph{runtime workload adaptivity} as the key factor that prevents existing libraries from achieving higher performance. This lack of adaptivity appears in two ways. First, existing work provides only a limited set of implementation variants for each database operator, restricting the choices available for different workloads. Second, existing libraries use little runtime information about intermediate data, such as value ranges, distinct counts, and string lengths, even though exactly this information determines which variant, configuration, or data representation is the most efficient. Knowing the number of distinct values (NDV) in a grouping column, for example, would let the engine pick the best group-by algorithm and size its hash table accordingly. The two aspects are complementary and both necessary for adaptivity: the first provides the set of available choices, while the second enables informed runtime decisions.

In this paper, we present Eiger, a GPU-based library designed to provide high-performance single-GPU query processing through runtime workload adaptivity. To this end, Eiger is designed around two principles.

\begin{enumerate}[wide, labelwidth=!, labelindent=0pt, topsep=0pt]
    \item \textbf{Multiple operator implementations.}
    Eiger provides multiple implementation variants for each operator, often with tunable knobs. For joins, Eiger implements non-partitioned hash, partitioned hash, and sort-merge algorithms, combinable with two payload materialization strategies. For group-bys, it provides hash-, sort-, and partition-based algorithms. For expression evaluation, it offers a per-tuple interpretation backend that keeps intermediate results in registers or shared memory, and a batch-based backend composed of type-specialized and vectorized kernels. For string matching, both the number of threads per string and the width of packed memory accesses are tunable, and substring matching can switch between a quadratic-time algorithm and the linear-time Knuth-Morris-Pratt algorithm. For sorting, Eiger chooses between radix sort and merge sort, the latter optionally accelerated by prefix extraction for strings. Table~\ref{tab:impl} summarizes all variants together with the workload properties that determine the best choice. As this catalog shows, Eiger practices the principle not only in traditionally expensive operators, such as joins and group-bys, but also in expensive operations that are often overlooked by the literature on GPU-accelerated databases. For the latter, Eiger contributes new techniques: packed (multi-byte) string accesses that handle arbitrary alignment, expression linearization based on the Sethi--Ullman ordering to minimize intermediate results, and an order-preserving dictionary encoding algorithm for GPUs.
    \item \textbf{Runtime adaptive execution.} Instead of relying solely on static query optimization, Eiger defers the choice of implementations and configuration parameters to \emph{query runtime} whenever possible. Unlike traditional CPU databases, where statistics are collected only for base tables beforehand, Eiger profiles intermediate data \emph{during query execution} using a curated set of lightweight statistics: minimum, maximum, and mean values, and a HyperLogLog++ sketch estimating the NDV. These statistics drive three kinds of runtime decisions: \emph{algorithm selection} (e.g., the estimated group cardinality selects the group-by algorithm and sizes its hash table), \emph{knob tuning} (e.g., string length statistics set the parallelism and packed access width for string matching), and \emph{on-the-fly data compression} (e.g., a mechanism we call \emph{smart key fusion} compresses multiple sort or group-by keys via frame-of-reference+bitpacking or order-preserving dictionary encoding and fuses them into a single 4- or 8-byte key, making the much faster radix sort applicable in place of merge sort). Our key insight is that GPUs compute such statistics at close to memory bandwidth, so profiling is cheap enough to easily pay off: computing an HLL++ sketch over $2^{28}$ keys takes about one millisecond, while choosing the wrong sort implementation in the same setting costs hundreds of milliseconds (Section~\ref{sec:eval-sort}).
\end{enumerate}

We evaluate Eiger with operator-level microbenchmarks and end-to-end queries on two different platforms (NVIDIA GH200 and A100); the results show that the best implementation depends not only on the workload but also on the GPU architecture. At the operator level, Eiger evaluates expressions up to 5$\times$ faster than cuDF, and smart key fusion accelerates multi-key sorting by up to 13$\times$ over merge sort. On the complete set of 22 TPC-H queries with scale factors 10, 30, and 100, Eiger outperforms a cuDF-based execution engine by up to 1.8$\times$ in total runtime and up to 6.1$\times$ on individual queries.

We summarize our key contributions as follows.
\begin{itemize}[wide, labelwidth=!, labelindent=0pt, topsep=0pt]
    \item We argue that runtime workload adaptivity, i.e., the combination of multiple implementation variants per operator and runtime statistics to choose among them, is the key ingredient to push single-GPU data analytics performance further, and we present Eiger, a high-performance library for GPU-based data analytics that demonstrates this idea.
    \item We propose optimization techniques for expensive but often neglected operations: packed accesses and cooperative-group-based parallelism for string matching, Sethi--Ullman linearization and register-resident intermediate results for expression evaluation, and smart key fusion with order-preserving dictionary encoding for multi-key sorting and group-bys.
    \item We introduce a runtime profiling and adaptive execution mechanism for GPU-based query execution, and show in three concrete scenarios (algorithm selection, knob tuning, and on-the-fly compression) how cheap GPU-side statistics translate into large performance gains.
    \item We conduct a comprehensive performance study with operator-level microbenchmarks on two GPU architectures and all 22 TPC-H queries at three scale factors, characterizing when each implementation variant wins and thereby providing a basis for cost models in future GPU query optimizers.
\end{itemize}

\section{Background}
\subsection{GPU Architecture and Programming}
\label{sec:background-gpu}
The primary execution unit of a GPU is the Streaming Multiprocessor (SM), which comprises control units, a large register file, CUDA cores, and on-chip fast memory, including the co-locating L1 cache/shared memory, and specialized caches such as the constant cache. Multiple SMs are connected via a unified L2 cache to off-chip global memory, which has higher latency and lower throughput. 

GPU programs are executed by a large number of lightweight threads, organized hierarchically into thread blocks and grids. Threads within a block can cooperate using shared memory and synchronization primitives, while blocks are scheduled independently across SMs. At the hardware level, threads are executed in fixed-size groups called warps, following a single-instruction, multiple-thread (SIMT) execution model. This organization enables efficient latency hiding by rapidly switching between ready warps.

From a programming perspective, GPU computation is expressed in terms of kernels, which define functions executed in parallel. A more recent programming abstraction, called \textit{cooperative group}, allows the programmer to program collaboration among threads more easily, no longer bound to the warp level or block level. 

\subsection{GPU-based Query Processing}
\label{sec:gpu-qp}
Query processing on the GPU either assumes the data (input, output, and any intermediate results) are completely resident in the GPU memory, or the data could move between the CPU memory, GPU memory, and sometimes even through a network. In the latter case, a buffer manager and/or a network manager is usually needed to administer the data movement. In this study, we focus on the former case because single-GPU in-memory processing serves as a foundation for all different scenarios. 

Unlike in CPU-based systems, which often implement a pipelined volcano or vector model, the most popular execution model for GPU-based query processing is full materialization. In other words, each operator waits for the last operator to completely finish before starting.
This is also the model assumed in this work and that of the most popular GPU-based operator library, cuDF.
The relations are usually stored in columns and almost never converted to a row format during execution. When an input relation has multiple columns, the operator often initializes a column of tuple IDs (TIDs) as the delegate of the columns that are not directly involved in the operation. Later, those columns will be materialized with the TIDs, which is also known as late materialization.

\section{Eiger Operators and Optimizations}
\label{sec:eiger-overview}
In this section, we explain operator implementations in Eiger. 
Table~\ref{tab:impl} summarizes them and lists the factors influencing when to use each. The factors are grouped into two categories: static ones, which can be known before the query execution, and runtime ones, which are only available during the query execution.

\subsection{Data Layout and Data Types}
\label{sec:data-layout}
Eiger adopts columnar storage, where each column of a relation is stored in one contiguous memory region, with the exception of strings (see Section~\ref{sec:strings}). Programming-wise, Eiger provides the typeless \texttt{column\_view} (immutable) and \texttt{mutable\_column\_view} (mutable) classes that can be passed directly as arguments to the device kernel without extra memory copy. 
A \texttt{(mutable\_)table\_view} is a collection of \texttt{(mutable\_)column\_view}, but cannot be passed to the device kernel. 
For GPU kernels that work on relations with arbitrary numbers of columns, we provide the \texttt{table\_device\_view}, which can be converted from the \texttt{table\_view} through very cheap memory copies. The \texttt{table\_device\_view} ensures that all meta information about the table is stored in the device global memory instead of in the thread's local memory. The latter may lead to substantial register pressure and consequently harm performance. 

Compared to cuDF, which uses the 32-bit integer as the ``size type'' to store the number of rows, Eiger adopts the 64-bit integer as the size type to be future-proof.
Our rationale is that, with the memory capacity of GPUs steadily increasing, the table size that can be handled can easily outgrow (or may have already outgrown) the 32-bit integer range. One example is that cuDF has already encountered the issue that the 32-bit integer is not enough to index all the bytes of a string column~\cite{cudf_issue_3958}.
However, moving from 32-bit to 64-bit size type is not free because handling 64-bit tuple IDs (TIDs) can be more expensive in some scenarios. For example, reading a 64-bit TID column results in 2 times more data being loaded from memory compared to a 32-bit TID column. This performance penalty is amplified when we need to load or store multiple passes of TIDs in an algorithm (e.g., sorting). Another example is that some implementations utilize shared memory for caching intermediate results, and using 64-bit TIDs may result in less data being cached, affecting the efficiency of the algorithm. 

\begin{table*}[t] % summary of algorithms
\centering
\caption{Summary of implementations in Eiger and factors to be considered for algorithm selection.}
\label{tab:impl}
\resizebox{\textwidth}{!}{%
\begin{threeparttable}
\begin{tabular}{|l|l|l|l|}
\hline
\textbf{Operation} &
  \textbf{Implementations/Techniques} &
  \textbf{Static workload properties} &
  \textbf{Runtime workload properties} \\ \hline
Join (\S \ref{sec:join}) &
  \begin{tabular}[c]{@{}l@{}}Sort-merge join\\ Partitioned hash join\\  Non-partitioned hash join\end{tabular} &
  \begin{tabular}[c]{@{}l@{}}Data types\\ Sorted-ness of join keys\end{tabular} &
  \begin{tabular}[c]{@{}l@{}}Cardinality of the relations$^\dagger$\\ Distribution of the join keys$^\ddagger$ \\ Output cardinality (match ratio)$^\ddagger$\end{tabular} \\ \hline
Materialize (\S \ref{sec:join}, \S \ref{sec:groupby})&
  \begin{tabular}[c]{@{}l@{}}Gather from untransformed relations (GFUR) \\ Gather from transformed relations (GFTR) \end{tabular} &
  Data types &
  \begin{tabular}[c]{@{}l@{}}Cardinality of gather map$^\dagger$\\ Distribution of gather map values$^\ddagger$\end{tabular} \\ \hline
Group-by (\S \ref{sec:groupby}) &
  \begin{tabular}[c]{@{}l@{}}Sort-based (optional dictionary encoding opt)\\ Partitioned hash-based\\ Non-partitioned hash-based\end{tabular} &
  \begin{tabular}[c]{@{}l@{}}Data types\\ Aggregation functions\end{tabular} &
  \begin{tabular}[c]{@{}l@{}}Cardinality of the relation$^\dagger$\\ Group cardinality (\S \ref{sec:ndv})$^\ddagger$\end{tabular} \\ \hline
String matching (\S \ref{sec:strings})&
  \begin{tabular}[c]{@{}l@{}}Packed access width\\ Varied parallelism\\ KMP algorithm for substring matching\end{tabular} &
  Length of the pattern string &
  \begin{tabular}[c]{@{}l@{}}Distribution of the string lengths$^\ddagger$\\ (\S \ref{sec:minmax})\end{tabular} \\ \hline
Expression eval (\S \ref{sec:expr-eval})&
  \begin{tabular}[c]{@{}l@{}}Per-tuple interpretation (reg/shared memory) \\ Batch-based processing\end{tabular} &
  \begin{tabular}[c]{@{}l@{}}Register usage\\ Intermediate results size\end{tabular} &
  Memory usage$^\dagger$ \\ \hline
Sort  (\S \ref{sec:sort}) &
  \begin{tabular}[c]{@{}l@{}}Radix sort\\ Smart key fusion (\S \ref{sec:skf})\\ Merge sort\\ (w/ prefix extraction for strings (\S \ref{sec:string-sort})) \end{tabular} &
  Data types of sort keys &
  \begin{tabular}[c]{@{}l@{}}Distribution of string prefixes (\S \ref{sec:ndv})$^\ddagger$\\ Key compressability (\S \ref{sec:skf})$^\ddagger$\end{tabular} \\ \hline
\end{tabular}%
\begin{tablenotes}
\item[$\dagger$] Readily accessible at the start of the operation;
\item[$^\ddagger$] Needs additional statistics about the data, such as $\min$, $\max$, ndv.
\end{tablenotes}
\end{threeparttable}
}
\end{table*}

\subsection{Joins}
\label{sec:join}
Eiger implements multiple join algorithms, 
including non-partitioned hash join, partitioned hash join, and sort-merge join.
For sort-merge join and partitioned hash join, 
Eiger adopts the implementations detailed in the work of Wu et al.~\cite{wu25-gpu-joins-groupby}.
For the hash join, Eiger chooses the same implementation as cuDF, 
which uses the \texttt{static\_multiset} provided by cuCollection~\cite{cuco} as the hash table implementation. 
The \texttt{static\_multiset} only stores the tuple IDs and uses the user-provided hash function 
and equality comparator to hash and compare the actual data. This has two advantages over using the \texttt{static\_multimap} in cuCollection, which stores key-value pairs.
First, using \texttt{static\_multiset} allows the hash join to work with arbitrary data types, such as strings,
and avoids being restricted to only 4 or 8-byte data types, currently a limitation of \texttt{static\_multimap}. 
Second, \texttt{static\_multimap} suffers from bad performance
when the key-value pair to insert cannot be packed into a compare-and-swap-supported data type. For example, an 8-byte key and 4-byte value cannot be packed into a 128-bit CAS-supported data type without extra padding. Using a \texttt{static\_multiset} avoids this issue by only inserting the TIDs, which is a single type.

Eiger includes two strategies proposed in previous work~\cite{wu25-gpu-joins-groupby} to materialize the payload columns of a join, 
that is, gather-from\allowbreak-untransformed-relations (GFUR) and gather-from-transformed-relations (GFTR). 
While GFTR only applies to the partitioned hash join and sort-merge join, GFUR is applicable to all three algorithms. The main difference between GFTR and GFUR is whether we let the payload columns go through the same transformation (sort/partition) as the join keys. 
For example, in the sort-merge join, GFTR sorts all the payload columns according to the key column so that the materialization can avoid expensive random accesses later. 

\subsection{Group-by}
\label{sec:groupby}
For group-by, Eiger provides three algorithms (sort-based, hash-based, and partition-hash-based) and the two ways of handling payload columns (GFUR and GFTR)~\cite{wu25-gpu-joins-groupby}.
Compared to previous work~\cite{wu25-gpu-joins-groupby}, Eiger introduces a few important improvements. First, Eiger uses the \texttt{static\_set} to implement the partitioned-hash-based group-by instead of the \texttt{static\_map} used in previous work, following a similar reasoning as for the join implementation. 
Second, Eiger extends the hash-based group-by to support multiple numeric and non-numeric grouping keys, often seen in TPC-H queries. Third, multiple grouping keys may be combined to improve efficiency (see Section~\ref{sec:skf}). Fourth, the partition-based group-by applies a hash function to the keys before partitioning to ensure more even partition sizes. 

\subsection{Strings}
\label{sec:strings}
String processing often shows up in queries, but has not been discussed in-depth in the context of GPU-accelerated databases. 
Existing work, which claims to support full TPC-H queries, often gets around string processing by converting strings into numerics using dictionary encoding~\cite{wu2025terabytescaleanalyticsblinkeye,mohr23-boss}. 
We argue that although dictionary encoding works well for low-cardinality string columns, processing uncompressed strings is still important because not all operations (e.g., substring matching) can work with dictionary-encoded numeric values. 

\subsubsection{Layout} 
In-memory string columns typically adopt one of two layouts, the old Arrow format~\cite{cudf-strings} and the German string format~\cite{umbra}. 
Like cuDF, Eiger chooses the Arrow format mainly for two reasons. 
First, the German string format needs more space to store the string column due to the explicit length field if the majority of strings cannot be inlined into the header. 
Second, depending on the length of the string, reading a German string may need three steps, which means the program often contains more branches. Our preliminary experimental study shows that the CUDA program for processing the German string format is often compiled to use more instructions than the Arrow format. Our study also shows that the German string format is only beneficial for prefix matching when strings are long and the pattern is fewer than 4 bytes.
On the other hand, the Arrow format only requires an offset array in addition to the string data, which results in only one additional lookup of the offset array. Because all the strings in a column are stored contiguously, it is also easy to exploit more optimization techniques (e.g., packed access), as we will show next. 

\begin{figure}[t]
    \centering
    \includegraphics[width=0.75\linewidth]{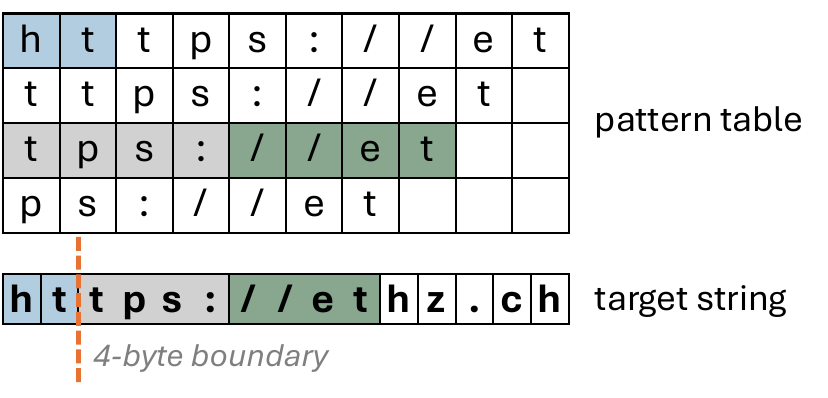}
    \caption{String matching with packed accesses.}
    \label{fig:str-match}
\end{figure}

\subsubsection{Packed access}
Accessing and processing one character (1 byte) at a time leads to high instruction counts and less efficient usage of memory bandwidth. 
To avoid this, whenever possible, Eiger uses a technique called packed access, which leverages wider load instructions (e.g., 4-byte) to load multiple characters at the same time and perform operations altogether (e.g., comparison). 
This reduces the number of instructions in the binary code and improves the efficiency of memory access. 

The challenge of enabling wider access is dealing with data alignment. Since strings are concatenated one-by-one in memory, the starting address of a string may not be aligned to be loaded as a wider data type. Even if we compare the unaligned part byte-by-byte, when the string finally becomes aligned, the pattern string may become unaligned at that byte. 
To solve this, if we want to use $k$-byte accesses, then we preprocess the pattern string by duplicating it $k$ times so that each character in the pattern string is aligned at $k$ bytes in one of the duplicates. All duplicates are stored in the shared memory of the thread block and can be created on the fly when reading the data from memory. During string comparison, when the beginning unaligned part of the target string is consumed, it then compares the rest with a duplicate of the pattern whose next byte to be compared is aligned. In this way, the rest of the comparison can be done with packed accesses except for a few trailing bytes. Figure~\ref{fig:str-match} illustrates how Eiger handles the alignment issue. The first two characters ``ht'' are compared byte-by-byte. Then, starting from the third character ``t'', we can compare 4 bytes at a time with the third duplicate of the pattern, whose third character is also aligned. 

\subsubsection{Cooperative groups}
As mentioned in Section~\ref{sec:background-gpu}, the cooperative group abstraction allows the programmer to flexibly assign a set of threads to a task. Eiger uses this to accelerate string matching operations.
Some pattern-matching operations, such as prefix and suffix matching, can benefit from more parallelism per string when the pattern string or target string is long. 
Moreover, when the target strings are also long, using more threads improves the cache locality.
Therefore, we utilize cooperative groups to vary the level of parallelism used to process each string. In contrast to the warp programming paradigm, where the whole warp (32 threads) is allocated for a task, using cooperative groups gives finer control over the parallelism (2, 4, 8 threads, and so on), depending on the size of the task. This technique can be 
combined with packed access to further improve performance.

\subsubsection{Substring matching}
Substring matching is commonly seen in queries. 
In addition to the vanilla quadratic-time string matching algorithm used in cuDF, Eiger also includes the Knuth-Morris-Pratt (KMP) algorithm~\cite{knuth1977-kmp} that works in linear time. The KMP algorithm has been shown to work well for GPUs~\cite{Branchini25-gpu-kmp}. It requires a preprocessing step on the pattern string to construct the prefix function. We let each thread block perform the preprocessing independently and store the prefix function in shared memory.

For the vanilla string matching algorithm, we can apply the packed access optimization; for the KMP implementation, we always use the byte-by-byte access.

\subsubsection{String sorting}
\label{sec:string-sort}
Since strings have variable lengths, linear-time radix sorting does not work. Instead, Eiger uses merge sorting with an important optimization for long strings. Before starting the merge sort, as an optional additional step, we extract the 4-byte prefix from all strings and store them contiguously in a prefix array. During the merge sort, the custom comparison functor will first compare the prefixes from the prefix array, and only when they have a tie, the functor will continue to check the actual strings. This method greatly accelerates the comparison by providing more efficient memory access. 
Note that extracting the prefixes will not always benefit the performance, especially when the prefixes are not discriminating enough. In Section~\ref{sec:runtime-exec}, we discuss how we decide if it is worth extracting prefixes during the query execution time. 

\subsection{Expression evaluation}\label{sec:expr-eval}
Expressions are an integral part of many database operators, most importantly, selection and projection. Recent work has pointed out that for a substantial number of queries on the TPC-H benchmark, the filtering operation in selections and projections is the most expensive part of the query when running on a GPU \cite{kabic25-maxbench}. 
Eiger has two expression evaluation backends: a per-tuple interpretation (PTI) backend and a batch-based (BB) backend. Both backends take as input a source table and an expression tree, which dictates how the expression should be evaluated to an output column. Before evaluation, the expression tree is traversed and linearized into a sequence of steps, each consisting of an operation, a column reference, or a constant. Unlike cuDF, which uses the post-order traversal to linearize the expression tree, Eiger adopts the Sethi-Ullman ordering~\cite{sethi-ullman}, which minimizes the peak number of intermediate results needed to evaluate the tree. Minimizing intermediate results benefits both backends, as we will explain soon. 

The PTI backend launches a single kernel that works as an expression tree interpreter to evaluate the expression for each tuple in the source table. Therefore, before launching the kernel, the linearized expression tree needs to be transferred to the device memory. The advantage of this backend is that it can prevent intermediate results from being written back to the device memory, which causes extra memory transactions. Despite the advantage, the challenge lies in storing and managing the intermediate results. A careless implementation may result in the intermediate results being stored in the thread-local memory, which is backed by the device memory under high register pressure. cuDF uses the shared memory for this purpose, where each intermediate result is written into a designated slot. The slot assignment can already be done during the expression tree linearization on the CPU. Eiger adopts a similar design, but additionally implements a specialization where the intermediate results are stored in the registers. When linearizing the expression tree, if the peak size of intermediate results is smaller than a threshold, we can store them in registers, which have an order of magnitude lower access latency than shared memory. Thanks to the Sethi-Ullman ordering, we use fewer registers (equivalently, intermediate results) to evaluate an expression compared to the post-order traversal. The disadvantage of the PTI backend is the high interpretation overhead. Due to the generality of the interpretation kernel, the decision of which operator to dispatch is made at runtime and for each tuple, leading to high instruction counts and pressure on the ALU pipeline in CUDA cores. 

Instead of relying on one generic interpreter kernel, the BB backend evaluates each step in the linearized expression tree using a separate kernel (except for column reference and constants). The benefit of this is that there is almost no interpretation overhead. Each kernel is specialized for the type of operation and its operands through C++ templates. All operations can be categorized into unary operations, binary operations, and ternary condition operations (e.g., \texttt{pred ? a : b}). Since the kernel always deals with one to three input columns and produces one output column, we can leverage the vectorized load and store to further improve memory bandwidth utilization and reduce overheads. The disadvantage of this backend is that it consumes extra device memory to store the intermediate results, and multiple kernels lead to more reads and writes into the device memory.

Choosing between PTI and BB depends heavily on the properties of the expression (e.g., intermediate result size), input relations (e.g., tuple size), and the GPU architecture (e.g., memory bandwidth vs. compute throughput). 

\subsection{Sorting}
\label{sec:sort}
Sorting is another often-neglected operator when studying the GPU-based databases, especially for the case of multi-key multi-data-type sorting. Many existing works also do not explain in detail how they handle sorting when it comes to multiple keys~\cite{mohr23-boss,wu2025terabytescaleanalyticsblinkeye,crystal_shanbhag_2020}. Eiger implements different ways to sort a table depending on the column data types.
Eiger prefers to use the radix sort whenever possible, due to its substantially better performance than the merge sort. Eiger uses merge sort when lexicographical comparisons are not possible, or the sorting is based on multiple keys.\footnote{Currently, the \texttt{cub::DeviceRadixSort} function in CUB only works for a single key.}
We categorize the sort implementations into the following cases.

\begin{itemize}[wide, labelwidth=!, labelindent=0pt, topsep=0pt]
    \item[\textbf{Case 1}] A single numerical sort key. We use the radix sort for the reason explained above.
    \item[\textbf{Case 2}] Multiple numerical sort keys. We will start with trying to combine the keys into one single key using the \emph{smart key fusion} (SKF) algorithm we propose in Section~\ref{sec:skf}. If combining them is possible, then we can process the rest like Case 1; otherwise, we will handle the sorting as Case 3 (see below).
    \item[\textbf{Case 3}] At least one non-numerical sort key. We use the merge sort with a custom comparator. The custom comparator maintains a \texttt{table\_device\_view} (see Section~\ref{sec:data-layout}) and sort orders (ascending or descending) for each sort key. Given two tuple IDs, the comparator retrieves the rows from the \texttt{table\_device\_view} and compares them based on the specified sort orders.
\end{itemize}

\section{Runtime Adaptive Execution}
\label{sec:runtime-exec}
In the previous section, we explained the implementations of Eiger operators in detail. A recurring pattern is that each implementation has its own favorable input characteristics for which it works most efficiently. If the query runtime can better understand input data, it can select a more efficient implementation or choose better configuration parameters. 
CPU-based databases often rely on statically maintained table or column statistics for this purpose~\cite{oracle-statistics,postgresql-statistics} and lack runtime profiling of the intermediate relations. 
For GPU-based databases, we argue that runtime profiling is beneficial in certain scenarios based on the following observations. First, many statistics, such as min, max, and mean, can be calculated extremely fast on the GPU (close to the memory bandwidth, which is currently a few TB/s). Although these statistics are basic, they offer valuable information about the data being handled. Second, choosing suboptimal implementations or configuration parameters is very costly, as evidenced in the experimental evaluation (Section~\ref{sec:evaluation}). Combining these two observations, as we will show in the experiments, the overhead of statistics computation for intermediate results is easily offset by the huge gain in choosing the optimal algorithm. Based on this, we propose that GPU-based query execution should embrace runtime profiling. In the rest of this section, we illustrate the idea with a few scenarios and show how calculating a curated set of statistics can boost the end-to-end operator performance. 

\subsection{Scenario I: Number of Distinct Values (NDV)}
\label{sec:ndv}
The performance of group-by is closely related to the group cardinality.
We propose using the lightweight and GPU-friendly HyperLogLog++~\cite{hyperloglogplusplus} (HLL++) algorithm to estimate this group cardinality before executing the group-by operator for large input sizes. The construction of the HLL++ sketch consumes almost negligible time compared to the actual group-by operator, but is helpful for identifying the optimal algorithm. In the case of hash-based group-by, if we estimate the group cardinality to be small, we can allocate a much smaller hash table to save memory space.
Similar ideas can be used to check whether the prefixes of a column of strings are diverse enough. This helps us decide if we should extract the prefixes to accelerate the sorting (see Section~\ref{sec:string-sort}). For strings that have common prefixes, such as HTTP websites, we avoid constructing the prefix array. 

\subsection{Scenario II: String Matching}
\label{sec:minmax}

The string matching algorithm in Eiger has two degrees of freedom, namely parallelism per string (except for substring matching) and packed access width.
To determine which configuration works better, we can calculate the \texttt{min}, \texttt{max}, and \texttt{mean} of the string lengths, which are stored as a separate array from the actual strings. 
With them, we can apply the following heuristic to pick the parameters: If both the average string length and the pattern length are long, we can increase both the parallelism and the packed access width. If only the average string length is long but the pattern is short, we should not use more parallelism, but can increase the packed access width. For the other cases, fixing the parallelism and access width to 1 is the most efficient.

\SetKwInput{KwInput}{Input}
\SetKwInput{KwOutput}{Output}
\newcommand\mycommfont[1]{\ttfamily\textcolor{blue}{#1}}
\SetCommentSty{mycommfont}
\SetKwComment{Comment}{$\triangleright$\ }{}

\begin{algorithm}[t]
\small
\caption{Order-preserving Dictionary Encoding}\label{alg:de}
\KwInput{Column $C_{in}$ to be encoded}
\KwOutput{Encoded column $C_{out}$, number of distinct values $ndv$}
\BlankLine

\SetKwFunction{FMain}{Main}
\SetKwProg{Fn}{Function}{:}{}
\Fn{\FMain{$C_{in}$}}{
    \Comment{(1) Assign group IDs and elect the group leader}
    $HT \gets \text{HashTable}(|C_{in}|)$\;
    $groupIdLoc \gets \text{alloc\_on\_device}(size\_t, |C_{in}|)$\;
    $ndv \gets 0$\;
    compute\_group\_id$(C_{in}, HT, groupIdLoc, ndv)$\;
    \Comment{(2) Retrieve key-ID pairs}
    [$uniqueKeys$, $groupIds$]$\gets HT.\text{retrieveAll()}$\;
    \Comment{(3) Argsort the unique keys to get order-preserving dictionary}
    $argsortIds \gets \text{argsort}(uniqueKeys)$\;
    \Comment{(4) Scatter the final encoded values}
    $sparseGroupIds \gets \text{alloc\_on\_device}(size\_t, |C_{in}|)$\;
    scatter$(groupIds, argsortIds, sparseGroupIds)$\;
    \Comment{(5) Encode all the keys}
    gather$(groupIdLoc, sparseGroupIds, C_{out})$\;
    \Return $C_{out},ndv$
}

\SetKwFunction{FCGI}{compute\_group\_id}
\Fn{\FCGI{$key$, $HT$, $gIdLoc$, $ndv$}}{
    \For{each thread $i$}{%
        \Comment{Insert or return the value if key has existed}
        $status, gIdLoc[i] \gets HT.\text{insert\_and\_find}(key[i],i)$\;
        \uIf{$status \text{ is Success}$} {
            $\text{atomicAdd}(\&ndv, 1)$\;
        }
    }
}

\SetKwFunction{FS}{scatter}
\Fn{\FS{$groupIds$, $argsortIds$, $sparseGroupIds$}}{
    \For{each thread $i$}{%
        $sparseGroupIds[groupIds[argsortIds[i]]] \gets i$\;
    }
}

\SetKwFunction{FG}{gather}
\Fn{\FG{$groupIdLoc$, $sparseGroupIds$, $C_{out}$}}{
    \For{each thread $i$}{%
        $C_{out}[i]\gets sparseGroupIds[groupIdLoc[i]]$\;
    }
}
\end{algorithm}

\subsection{Scenario III: Smart Keys Fusion}
\label{sec:skf}
Databases often need to handle multiple keys in operators like join, group-by, and sorting.
Managing multiple keys efficiently on the GPU is particularly challenging. 
The first reason is that each comparison (\texttt{==} or \texttt{<}) requires reading multiple keys, which are not stored together in the memory due to the columnar layout.
The second reason is that it is harder to implement efficient algorithms for multiple keys due to the limited shared memory capacity. 

To solve this, we propose a smart key fusion mechanism (SKF) that can assess the compressibility of each key and combine them into a single key after compressing them. It works like the following.
At \emph{query execution time}, for each key, SKF will evaluate the range of values by finding the minimum and maximum. If the range ($\max-\min$) is narrower than a preset threshold, we will compress the key by frame-of-reference encoding and bitpacking. Otherwise, we will try to calculate the HLL++ sketch on the key and find the number of distinct values (NDV). If NDV is below a certain threshold, we will use dictionary encoding to compress the keys. After the compression, we are able to represent each key with fewer bits. Then, we will invoke a single kernel to fuse the compressed keys by bit shifting into 4-byte integers or 8-byte integers. During fusion, we will alter the keys based on the data type (e.g., signed vs. unsigned) and sorting order requirements to ensure correctness.

Out of the above process, when used for sorting, the dictionary encoding is particularly tricky because it must preserve the order of the keys in the encoded values for the obvious correctness reason. To achieve this, we propose an \emph{order-preserving} dictionary encoding algorithm for the GPU (Algorithm~\ref{alg:de}).
In step 1, for each unique key, a leader tuple is picked, and the followers record the leader tuple ID in their own entry in the $groupIdLoc$ array (function \texttt{compute\_group\_id}). The function also returns the number of distinct values $ndv$. 
In step 2, we retrieve the pairs $(uniqueKey,groupId)$ from the hash map (line 6). There are in total $ndv$ pairs. The $groupId$s record the tuple IDs of leader tuples.
In step 3, we argsort the unique keys retrieved from the last step (line 7). The $argsortIds$ tells which position a unique key stays in before the argsort.
In step 4, we initialize the $sparseGroupIds$ array with $|C_{in}|$ elements and then fill in the entry of each group leader with their order-preserving encoded value (lines 8-9).
Lastly, in step 5, all items $C_{in}$ fetch the encoded value stored in their leader's entry in $sparseGroupIds$ (line 10). 

\section{Experimental Evaluation}
\label{sec:evaluation}

This section evaluates Eiger using microbenchmarks for 
joins, group-by, expression evaluation, string processing, and sorting, as well as end-to-end with the queries of the TPC-H benchmark.

\subsection{Experimental Setup}
\label{sec:eval-setup}

\paragraph{Hardware and system software.}
We use two GPU platforms:

\begin{itemize}[wide, labelwidth=!, topsep=0pt]
    \item \textbf{A100.} 
    NVIDIA A100-SXM4 (40\,GB HBM2e memory). 
    CUDA~12.4 and NVIDIA driver~550.90.07. 
    The host system is x86\_64 with an Intel Xeon CPU (12 cores; 2 sockets; SMT enabled).
    This is provided as the \texttt{a2-highgpu-1g} machine type from Google Cloud.
    \item \textbf{GH200.} 
    NVIDIA GH200 (96\,GB HBM3 memory). 
    CUDA~13.1 and NVIDIA driver~550.54.15. 
    The host system is aarch64 with an Arm Neoverse-V2 CPU (288 cores; 4 sockets).
    This is provided by the Swiss National Supercomputing Center.
\end{itemize}

If the results from both platforms show similar conclusions, we will only include the results from GH200. 

\paragraph{Baselines and methodology.}
Our baseline is cuDF v25.12. 
Unless otherwise stated, all benchmarks operate on GPU-resident data and report operator throughput as measured by our benchmark harness. 
We use the \texttt{nvbench} benchmark framework for our methods, which repeats the runs until the measurement has a noise level below 0.5\%. 
For cuDF measurements, we use their \texttt{pylibcudf} for better programmability without introducing extra Python-incurred overhead. In addition, we use a pre-allocated memory pool to reduce allocator-induced instability during the experiments, 
and we report the minimum time over repeated runs to be conservative.
We additionally force deallocation using \texttt{del} and \texttt{gc.collect()} in the benchmark scripts to avoid OOM errors during sweeps. 

For microbenchmarks that involve numeric values, we measure both 4-byte data types (I32) and 8-byte data types (I64).

\subsection{Joins}
\label{sec:eval-join}

\begin{figure}[t] % vary left narrow
    \centering
    \begin{subfigure}[b]{0.48\columnwidth}
        \centering
        \includegraphics[width=\columnwidth]{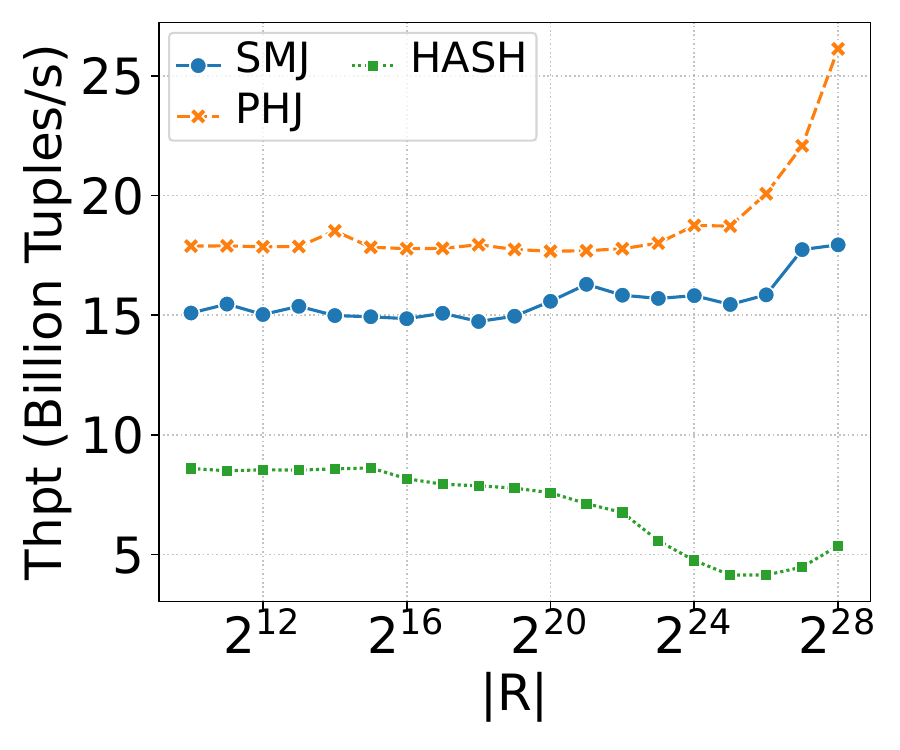}
        \caption{4-byte join key.}
        \label{fig:vary-right-narrow-join-4b}
    \end{subfigure}
    \hfill
    \begin{subfigure}[b]{0.48\columnwidth}
        \centering
        \includegraphics[width=\columnwidth]{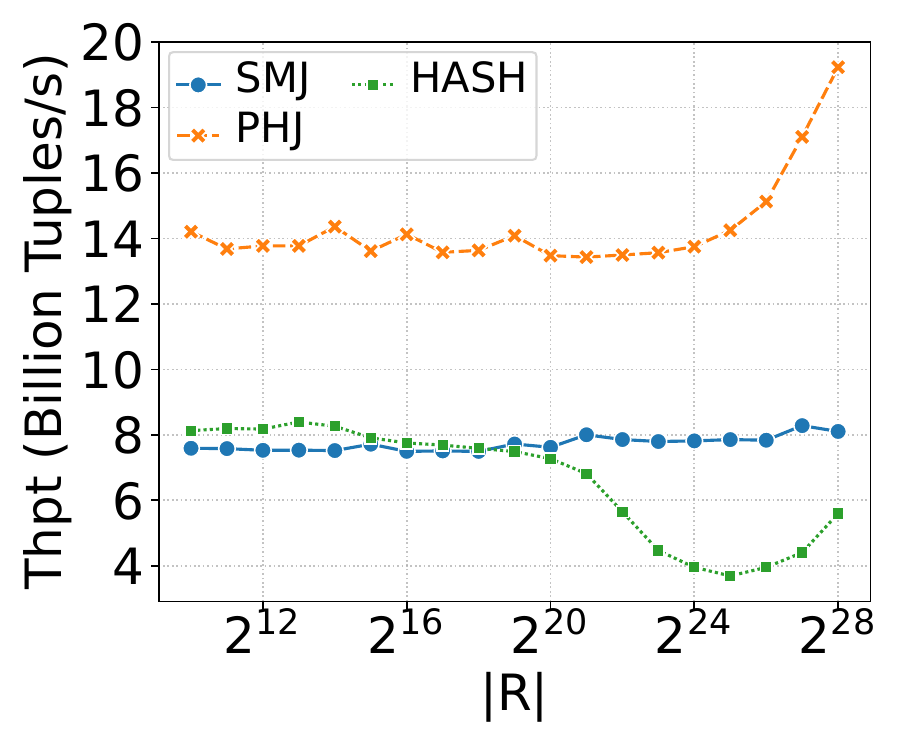}
        \caption{8-byte join key.}
        \label{fig:vary-right-narrow-join-8b}
    \end{subfigure}
    \caption{Narrow join with fixed $|S|=2^{28}$ (GH200).}
    \label{fig:vary-right-narrow-join}
\end{figure}

\begin{figure}[t] % vary right wide
    \centering
    \begin{subfigure}[b]{0.48\columnwidth}
        \centering
        \includegraphics[width=\columnwidth]{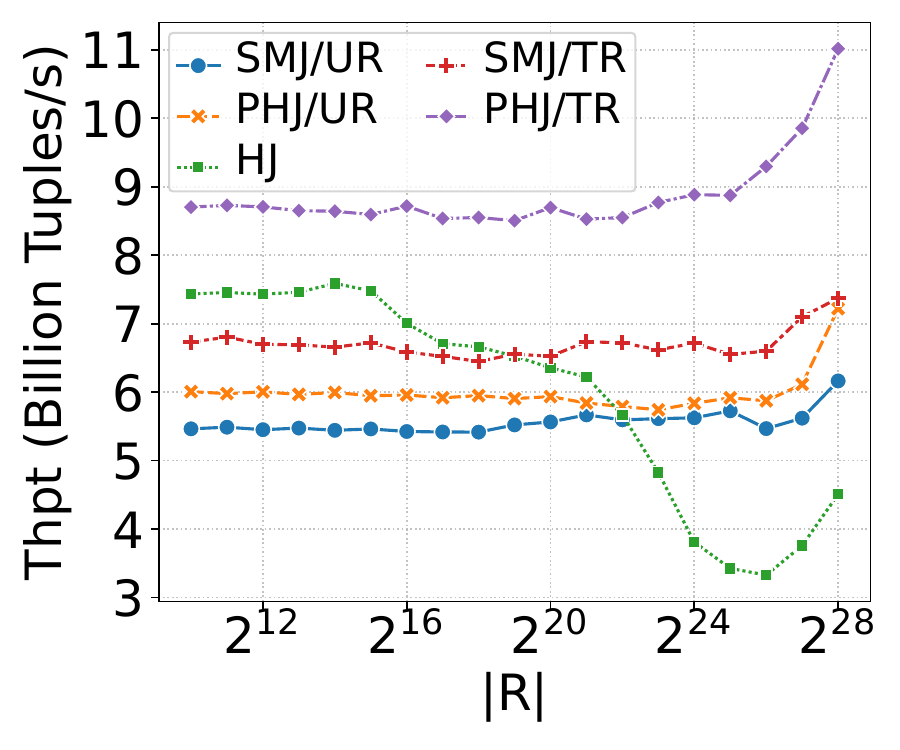}
        \caption{4-byte join key (GH200).}
        \label{fig:vary-left-wide-join-4b-GH200}
    \end{subfigure}
    \hfill
    \begin{subfigure}[b]{0.48\columnwidth}
        \centering
        \includegraphics[width=\columnwidth]{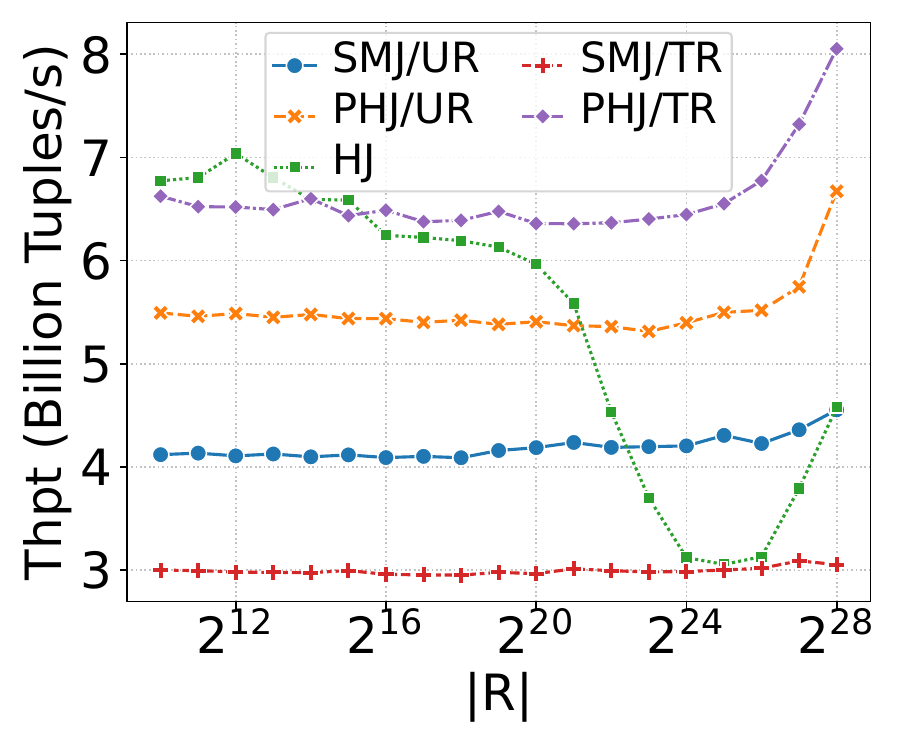}
        \caption{8-byte join key (GH200).}
        \label{fig:vary-left-wide-join-8b-GH200}
    \end{subfigure}
    \hfill
    \begin{subfigure}[b]{0.48\columnwidth}
        \centering
        \includegraphics[width=\columnwidth]{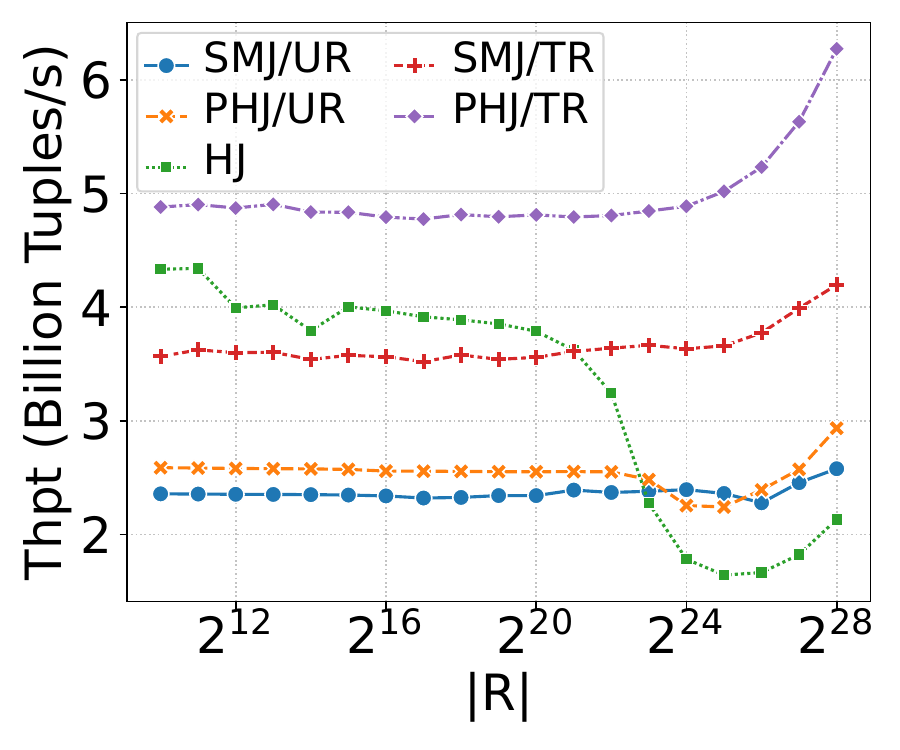}
        \caption{4-byte join key (A100).}
        \label{fig:vary-left-wide-join-4b-A100}
    \end{subfigure}
    \hfill
    \begin{subfigure}[b]{0.48\columnwidth}
        \centering
        \includegraphics[width=\columnwidth]{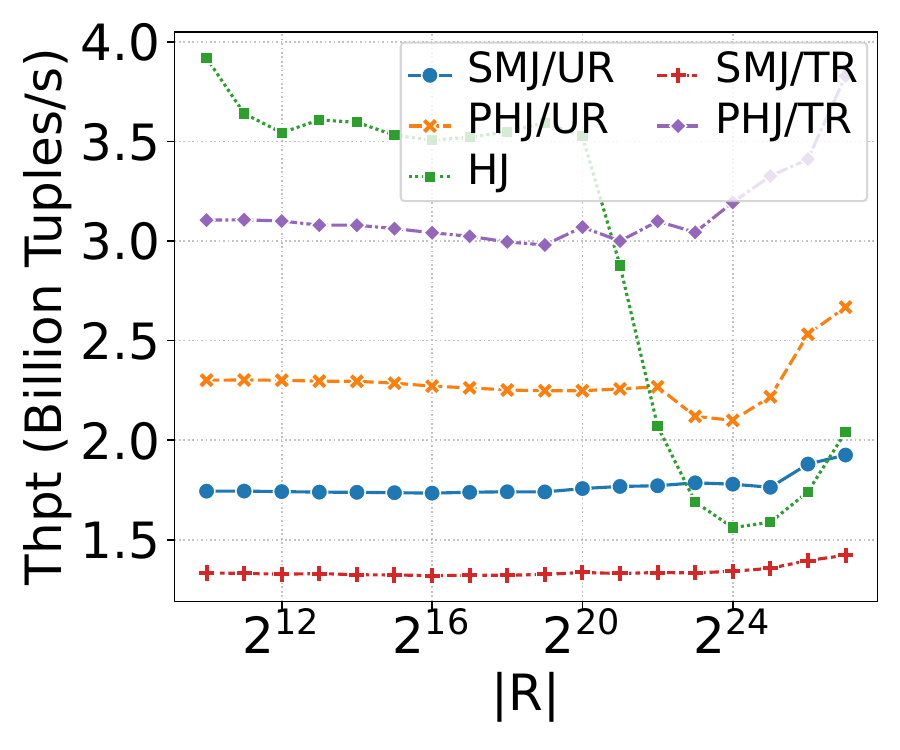}
        \caption{8-byte join key (A100).}
        \label{fig:vary-left-wide-join-8b-A100}
    \end{subfigure}
    \caption{Wide join with fixed $|S|=2^{28}$ (\ref{fig:vary-left-wide-join-4b-GH200}-\ref{fig:vary-left-wide-join-4b-A100}) or $|S|=2^{27}$ (\ref{fig:vary-left-wide-join-8b-A100}).}
    \label{fig:vary-left-wide-join}
\end{figure}

We evaluate the hash join (HJ), partitioned hash join (PHJ), and sort-merge join (SMJ). Our hash join uses the same \texttt{static\_multiset} implementation from cuCollection with cuDF; therefore, we skip comparing to cuDF in this experiment. For PHJ and SMJ, we consider two materialization strategies: GFUR (gather-from-untransformed-relations) and GFTR (gather-from-transformed-relations). HJ is only compatible with GFUR. When both relations have at most two columns, GFUR and GFTR are equivalent. For brevity, we use the notation ``SMJ/UR'' to indicate the ``join algorithm/materialization strategy'' in the text.

We denote the smaller side of the join as relation $R$ and the larger side as relation $S$. The size of the relation is indicated by $|\cdot|$. The throughput of the join is defined as $(|R|+|S|)/\text{Time}$.

\subsubsection{Varying the Left Table Size}
\label{sec:join-left-size}

In this experiment, we fix $|S|=2^{28}$ and vary $|R|$, the size of the smaller relation $R$. This is to simulate a typical type of join, where at least one side of the join is very large in number of rows. Depending on whether we need to materialize the payloads, we study the narrow join (no materialization) and the wide join (needs materialization).

\paragraph{Narrow join.} In this case, each relation has two columns and materialization of non-key columns can be ``inlined'' into the join, so no extra materialization is needed.

Figure~\ref{fig:vary-right-narrow-join} shows that with 4-byte keys and payloads, both SMJ and PHJ significantly outperform HJ across all left table sizes. This comes as a surprise because HJ is commonly considered very efficient for joins with a small build relation. The reason for the inferior performance is that HJ, in this case, inserts and probes with \emph{8-byte} tuple IDs (see Section~\ref{sec:data-layout} and Section~\ref{sec:join}), while SMJ and PHJ process the 4-byte keys and payloads without using TIDs. Compared to SMJ, PHJ is consistently faster. 
In general, the throughput of PHJ and SMJ increases with larger $|R|$, while the throughput of HJ decreases. 

For 8-byte keys and payloads, SMJ performance decreases substantially compared to the 4-byte case and becomes close to HJ for small-to-medium $|R|$; however, SMJ still outperforms HJ at large left table sizes.
PHJ performance remains the best across all $|R|$s. This is surprising because even though both HJ and PHJ deal with 8-byte data types, HJ still does not have an advantage over PHJ for small-to-medium $|R|$s. 

\paragraph{Wide join.}
In this case, each relation has four columns, and materialization of three non-key columns is necessary.

Figure~\ref{fig:vary-left-wide-join} repeats the experiment with four columns per table (two additional payload columns per relation compared to the narrow join). Here, A100 and GH200 show different behaviors. 

For 4-byte keys, PHJ/TR achieves the best performance across all configurations for both GH200 and A100 GPUs. HJ is second-best only when the left table is sufficiently small; this threshold differs between platforms: HJ remains competitive up to $|R|\le 2^{22}$ on A100, but only up to $|R|\le 2^{19}$ on GH200, after which it quickly falls behind. 
Compared to the narrow join, HJ no longer has the worst performance for small-to-medium $|R|$ due to its more efficient materialization. When the build relation is small, materializing the payloads from it is relatively efficient, and materializing the payloads from the probe side always involves sequential accesses. Across all sizes, GFUR-based approaches are consistently slower than their GFTR counterparts. 

For 8-byte keys, on GH200, HJ performs best for small left tables and is then overtaken by PHJ/TR as the left table grows; on A100, HJ remains the best until $|R|=2^{20}$. For both platforms, PHJ remains faster than SMJ regardless of the materialization strategy, and SMJ/TR is the slowest algorithm.

\begin{figure}[t] % Join: vary-right
    \centering
    \begin{subfigure}[b]{0.48\columnwidth}
        \centering
        \includegraphics[width=\columnwidth]{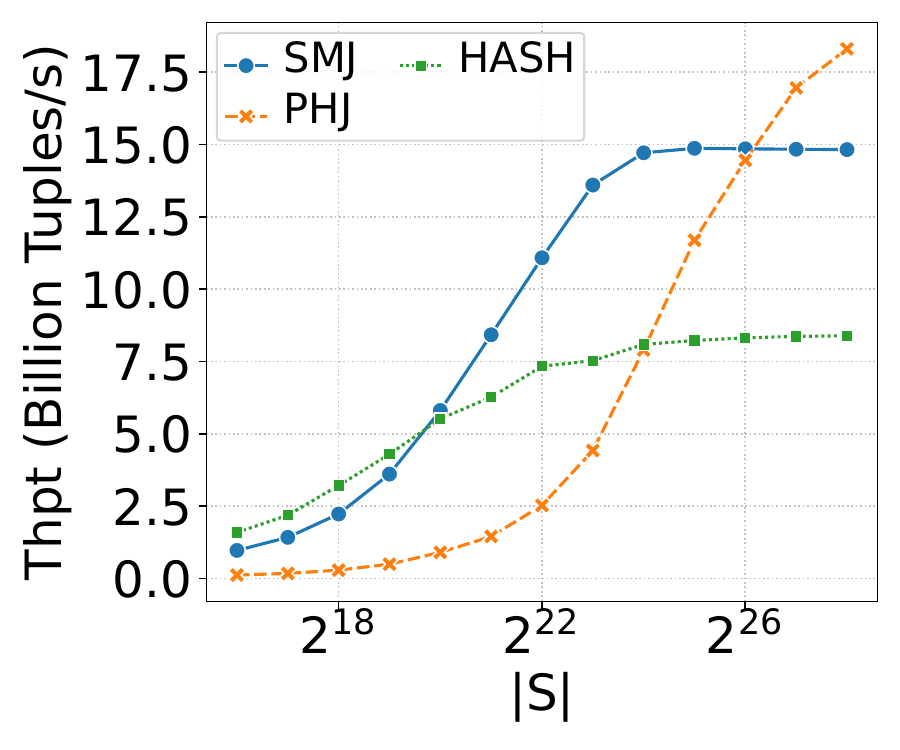}
        \caption{Narrow join: 4-byte join key.}
        \label{fig:vary-right-narrow-join-4b-GH200}
    \end{subfigure}
    \hfill
    \begin{subfigure}[b]{0.48\columnwidth}
        \centering
        \includegraphics[width=\columnwidth]{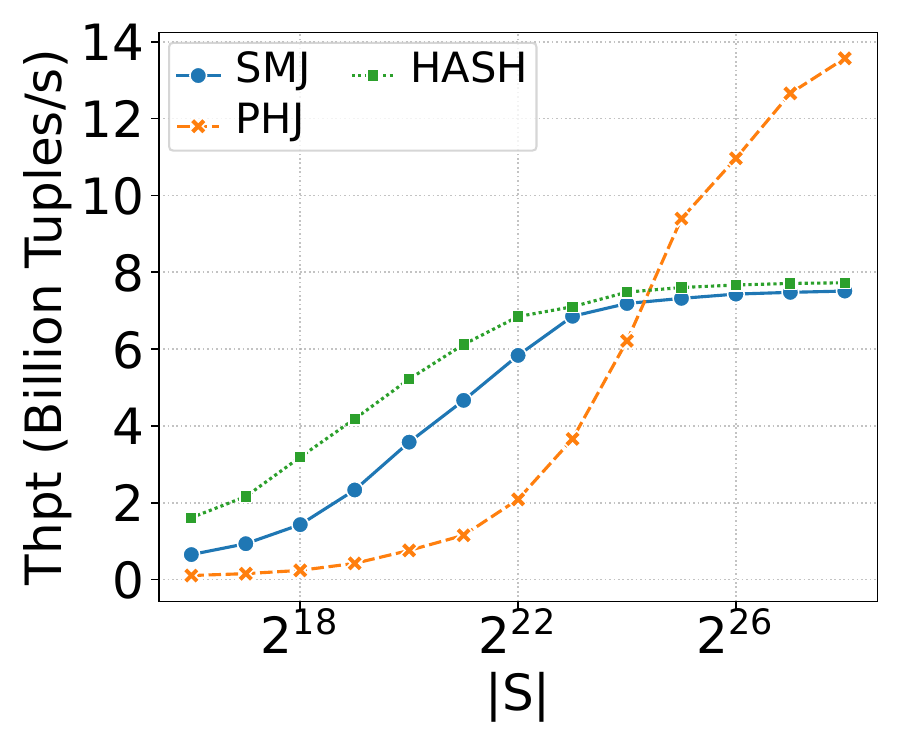}
        \caption{Narrow join: 8-byte join key.}
        \label{fig:vary-right-narrow-join-8b-GH200}
    \end{subfigure}
    \hfill
    \begin{subfigure}[b]{0.48\columnwidth}
        \centering
        \includegraphics[width=\columnwidth]{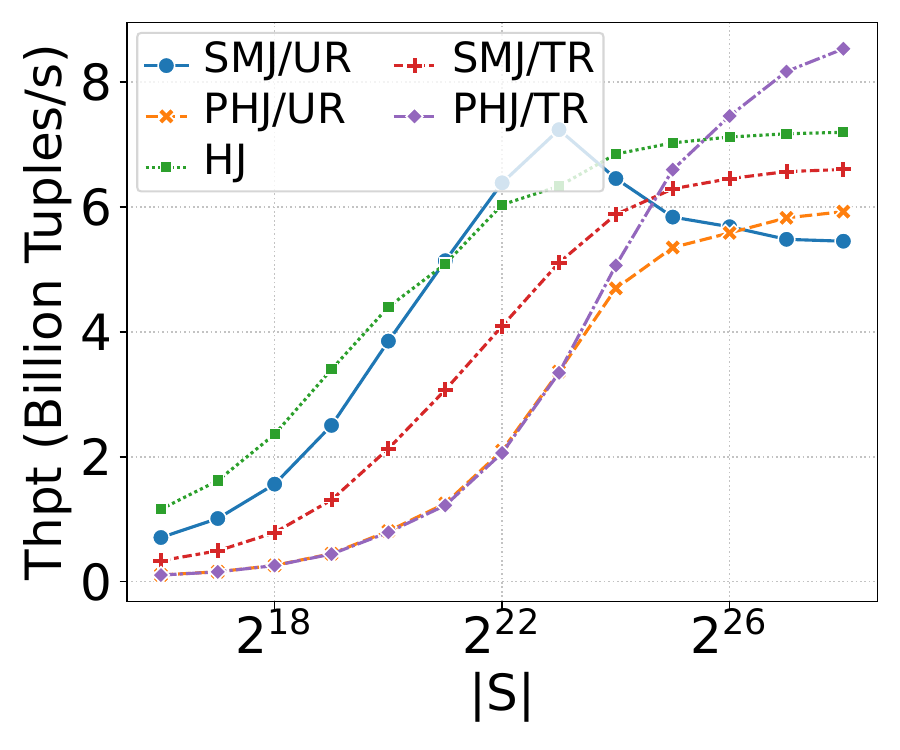}
        \caption{Wide join: 4-byte join key.}
        \label{fig:vary-right-wide-join-4b-GH200}
    \end{subfigure}
    \hfill
    \begin{subfigure}[b]{0.48\columnwidth}
        \centering
        \includegraphics[width=\columnwidth]{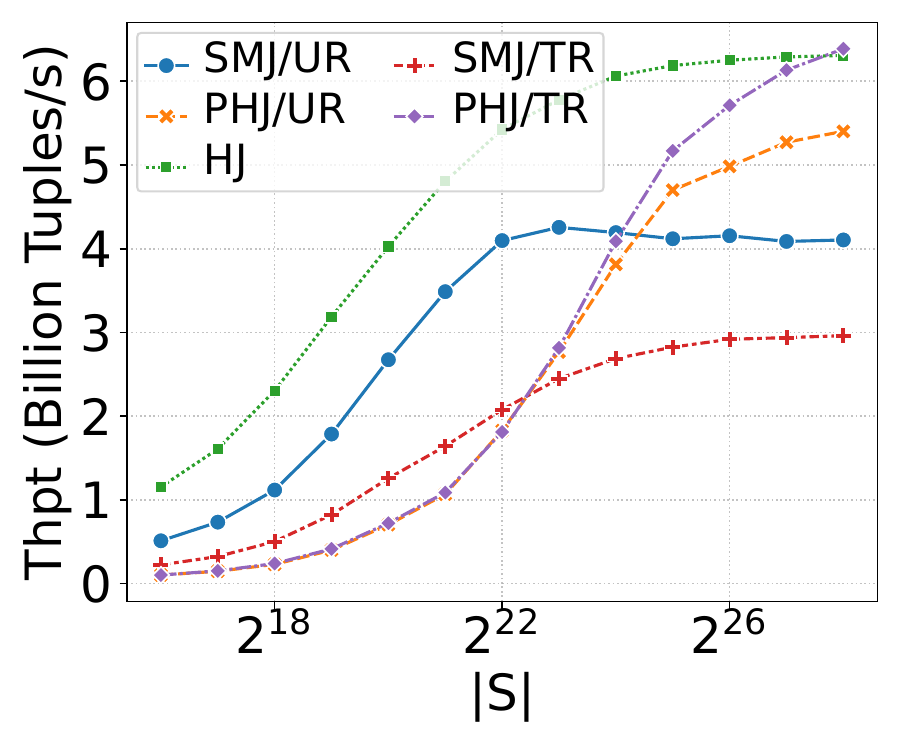}
        \caption{Wide join: 8-byte join key.}
        \label{fig:vary-right-wide-join-8b-GH200}
    \end{subfigure}
    \caption{Vary $|S|$ with fixed $|R|=2^{16}$ (GH200).}
    \label{fig:vary-right-join}
\end{figure}

\subsubsection{Varying the Right Table Size}
\label{sec:join-right-size}

In the following set of experiments, we do the opposite of the previous experiment by fixing $|R|=2^{16}$ and varying the size of $|S|$ while keeping $|S| \ge |R|$. This allows us to understand medium-sized joins better, where neither of the two relations is too large. 

Figure~\ref{fig:vary-right-join} shows the results for GH200. A100 has similar results.
In the narrow join scenario, for both 4-byte and 8-byte join keys, HJ performs the best for small $|S|$ (up to $2^{19}$ for 4-byte keys, and up to $2^{22}$ for 8-byte keys). This agrees with the results of the previous experiment that HJ has a greater advantage for I64 keys. 
Interestingly, the results show that SMJ outperforms PHJ for medium $S$ ($\le 2^{26}$ for I32 and $\le 2^{24}$ for I64). This is in contrast with the results from the last section, where PHJ is always better than SMJ for a very large relation $S$.

In the wide join scenario, the HJ has the best or second-best performance across $|S|$ for I32 and consistently outperforms other methods for I64. The advantage of SMJ over PHJ remains in the wide join for small-to-medium-sized relation $S$. For 4-byte keys, SMJ/UR is more efficient than SMJ/TR for $|S| \le 2^{24}$ but is less efficient for larger $|S|$s. For 8-byte keys, SMJ/UR is always better than SMJ/TR. PHJ/TR and PHJ/UR, for both I32 and I64, start with similar performance, but PHJ/TR performs better as $|S|$ grows.

This experiment reveals how different join algorithms perform under medium problem sizes (roughly measured by $|R|+|S|$), whereas previous work~\cite{wu25-gpu-joins-groupby} often focuses on large problem sizes. We demonstrate that in this case, the relative performance of different algorithms is significantly different from the large-sized joins. 

\subsubsection{Varying the Match Ratio}
\label{sec:join-match-ratio}

\begin{figure}[t] % match ratio
    \centering
    \begin{subfigure}[b]{0.48\columnwidth}
        \centering
        \includegraphics[width=\columnwidth]{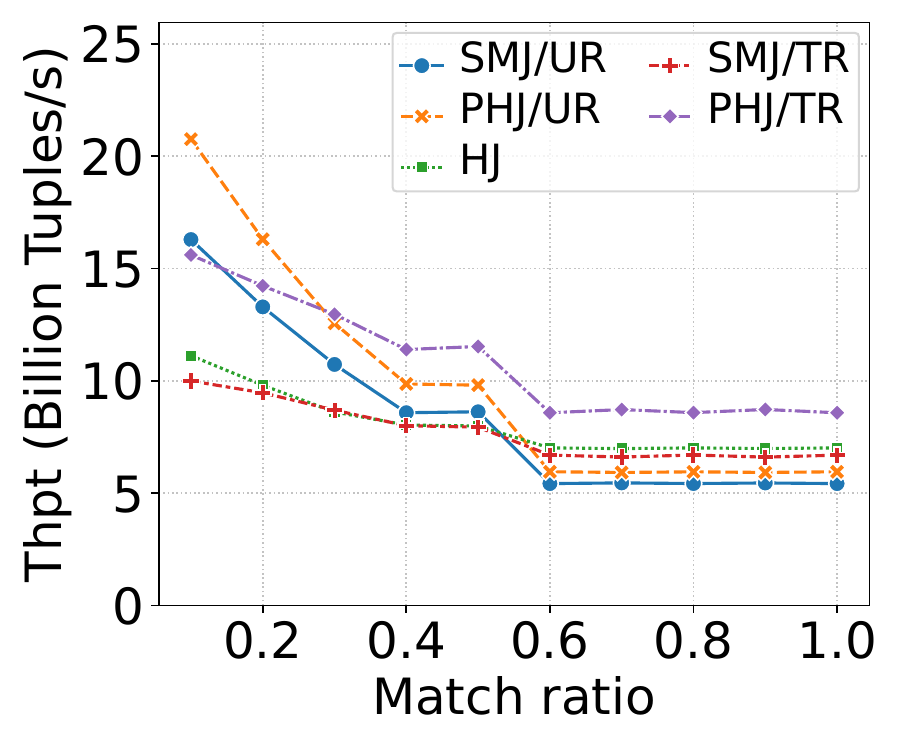}
        \caption{4-byte join key (GH200).}
        \label{fig:match-ratio-4b-GH200}
    \end{subfigure}
    \hfill
    \begin{subfigure}[b]{0.48\columnwidth}
        \centering
        \includegraphics[width=\columnwidth]{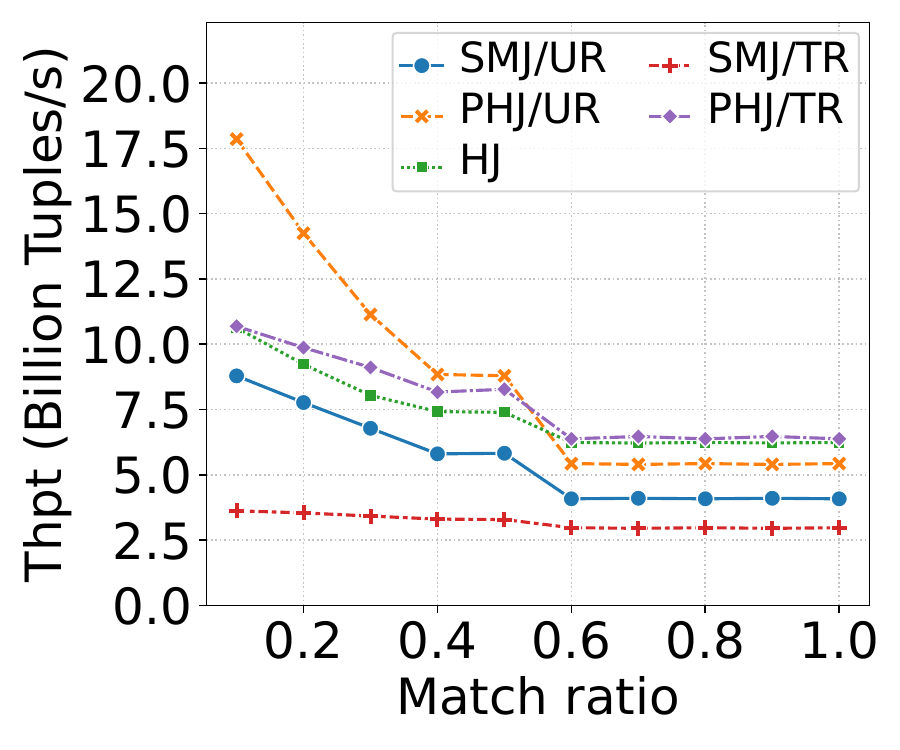}
        \caption{8-byte join key (GH200).}
        \label{fig:match-ratio-8b-GH200}
    \end{subfigure}
    \hfill
    \begin{subfigure}[b]{0.48\columnwidth}
        \centering
        \includegraphics[width=\columnwidth]{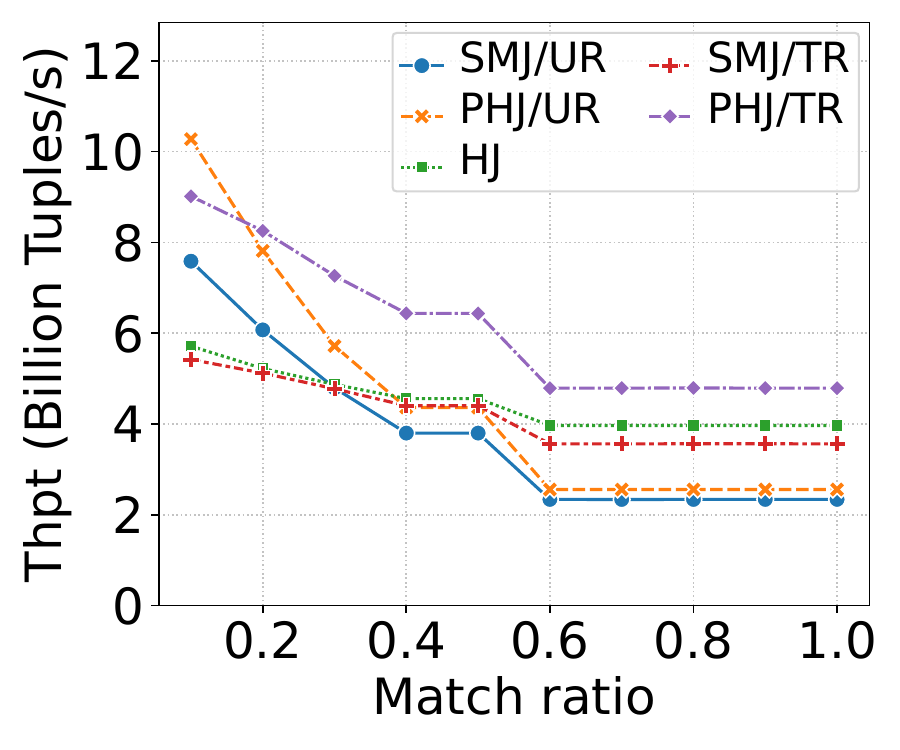}
        \caption{4-byte join key (A100).}
        \label{fig:match-ratio-4b-A100}
    \end{subfigure}
    \hfill
    \begin{subfigure}[b]{0.48\columnwidth}
        \centering
        \includegraphics[width=\columnwidth]{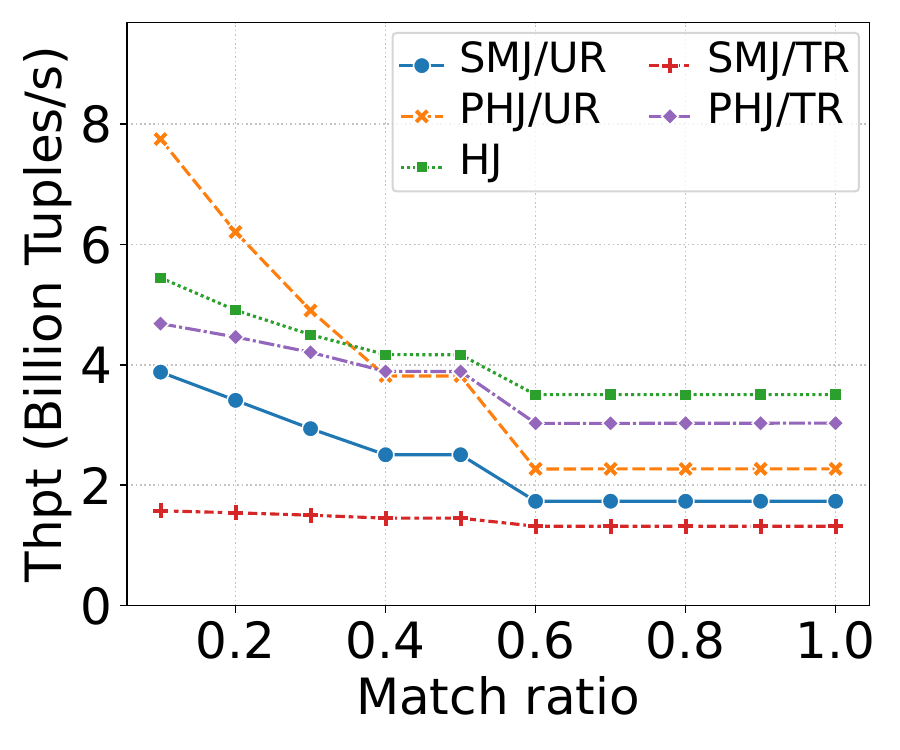}
        \caption{8-byte join key (A100).}
        \label{fig:match-ratio-8b-A100}
    \end{subfigure}
    \caption{Effect of match ratio on join performance.}
    \label{fig:match-ratio}
\end{figure}

In this set of experiments, we study the effect of the join match ratio, which is defined as the percentage of tuples from relation $S$ that have a match in relation $R$. The match ratio can influence the efficiency of materialization. 
Due to page limitations, we only show the subset of results where $|S|=2^{28}$. The only exception is Figure~\ref{fig:match-ratio-8b-A100}, where we set $|S|=2^{27}$ to avoid running out of memory.
Previous work~\cite{wu25-gpu-joins-groupby} has studied in-depth the case where both $R$ and $S$ are large; therefore, here we study a medium-sized $R$, with $|R|=2^{16}$. Each table has four columns. 

Figure~\ref{fig:match-ratio} shows the results of this set of experiments. For 4-byte join keys, when the match ratio is lower than a certain threshold (30\% for GH200 and 20\% for A100), PHJ/UR is the most efficient because the random gathering cost is low. Beyond this threshold, PHJ/TR is the most efficient, followed by HJ and SMJ/TR. 
For 8-byte keys, PHJ/UR remains the best algorithm for higher match ratios on both platforms before being overtaken by PHJ/TR and HJ. 

When comparing the GH200 and A100 GPUs, we find that on GH200, a higher match ratio (hence a higher materialization cost) is needed to justify the usage of the GFTR technique. This could imply that the random memory access performance is improved from A100 to GH200. 

\subsection{Group-By}
\label{sec:eval-groupby}

\begin{figure}[t] % group-by
    \centering
    \begin{subfigure}[b]{0.48\columnwidth}
        \centering
        \includegraphics[width=\columnwidth]{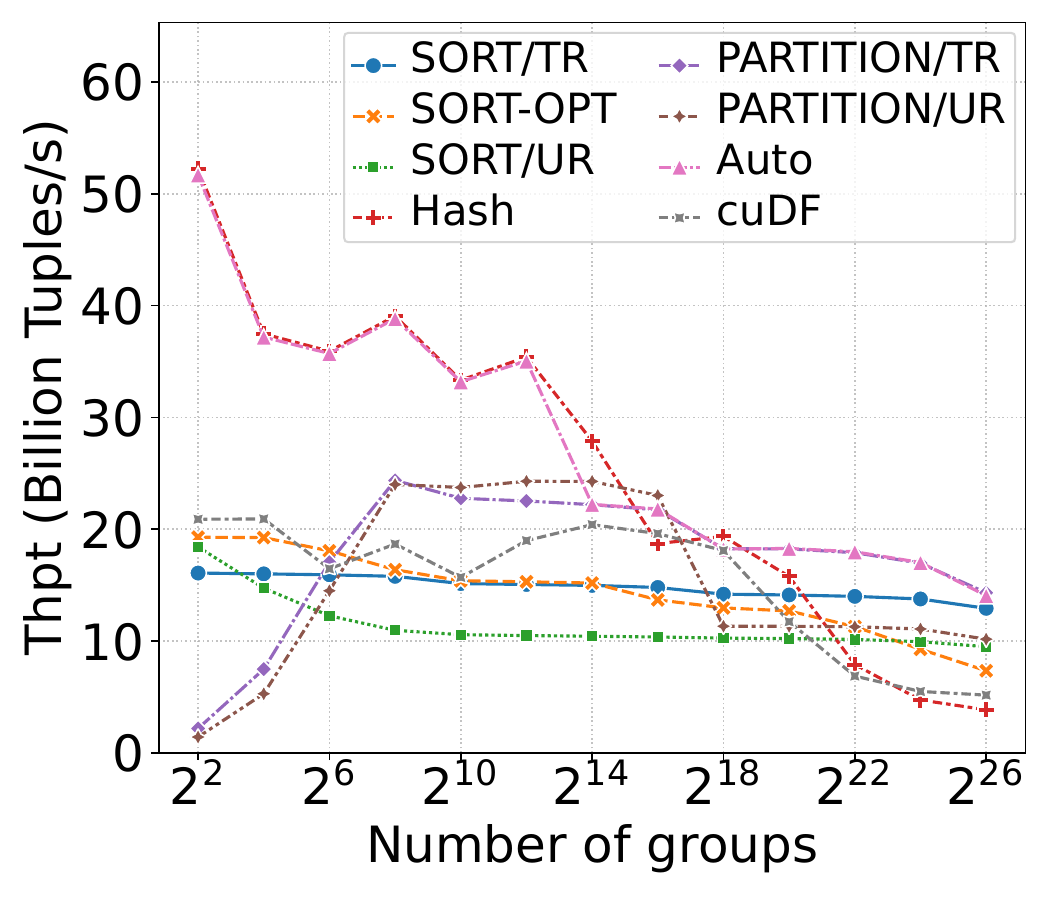}
        \caption{4-byte group key (GH200).}
        \label{fig:groupby-4b-GH200}
    \end{subfigure}
    \hfill
    \begin{subfigure}[b]{0.48\columnwidth}
        \centering
        \includegraphics[width=\columnwidth]{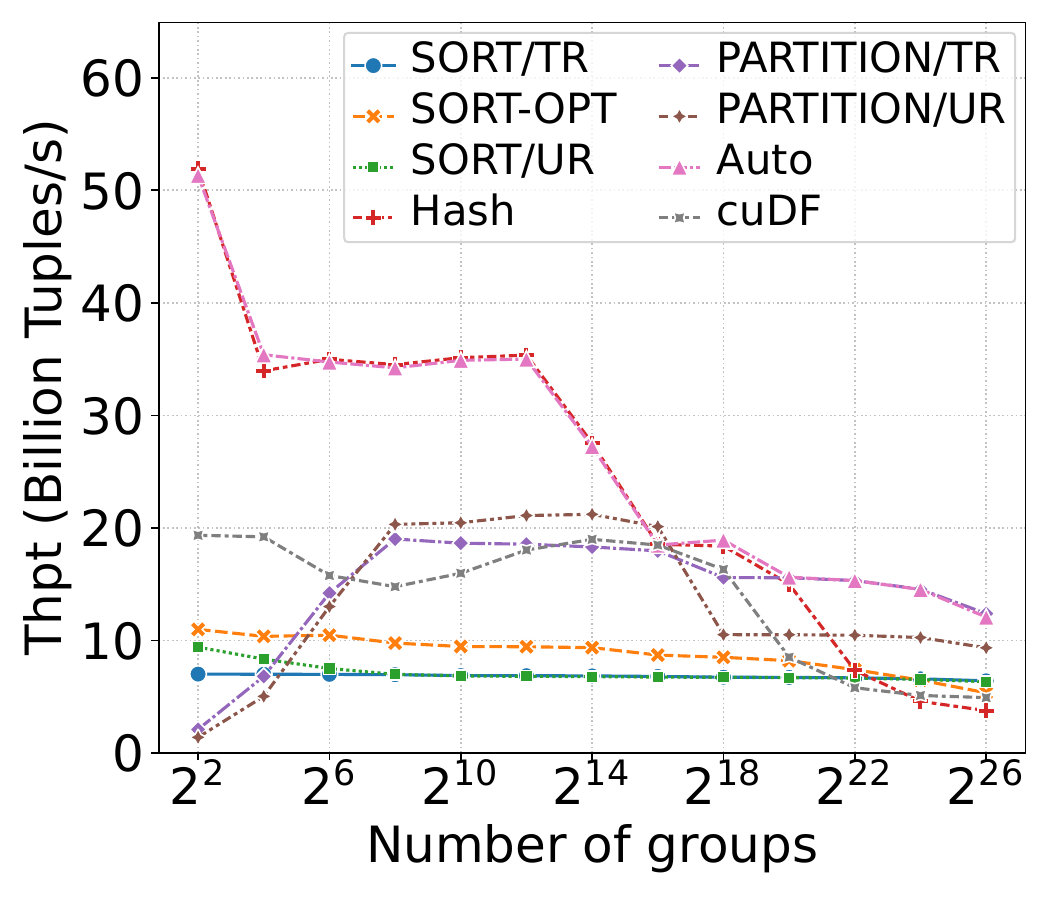}
        \caption{8-byte group key (GH200).}
        \label{fig:groupby-8b-GH200}
    \end{subfigure}
    \hfill
    \begin{subfigure}[b]{0.48\columnwidth}
        \centering
        \includegraphics[width=\columnwidth]{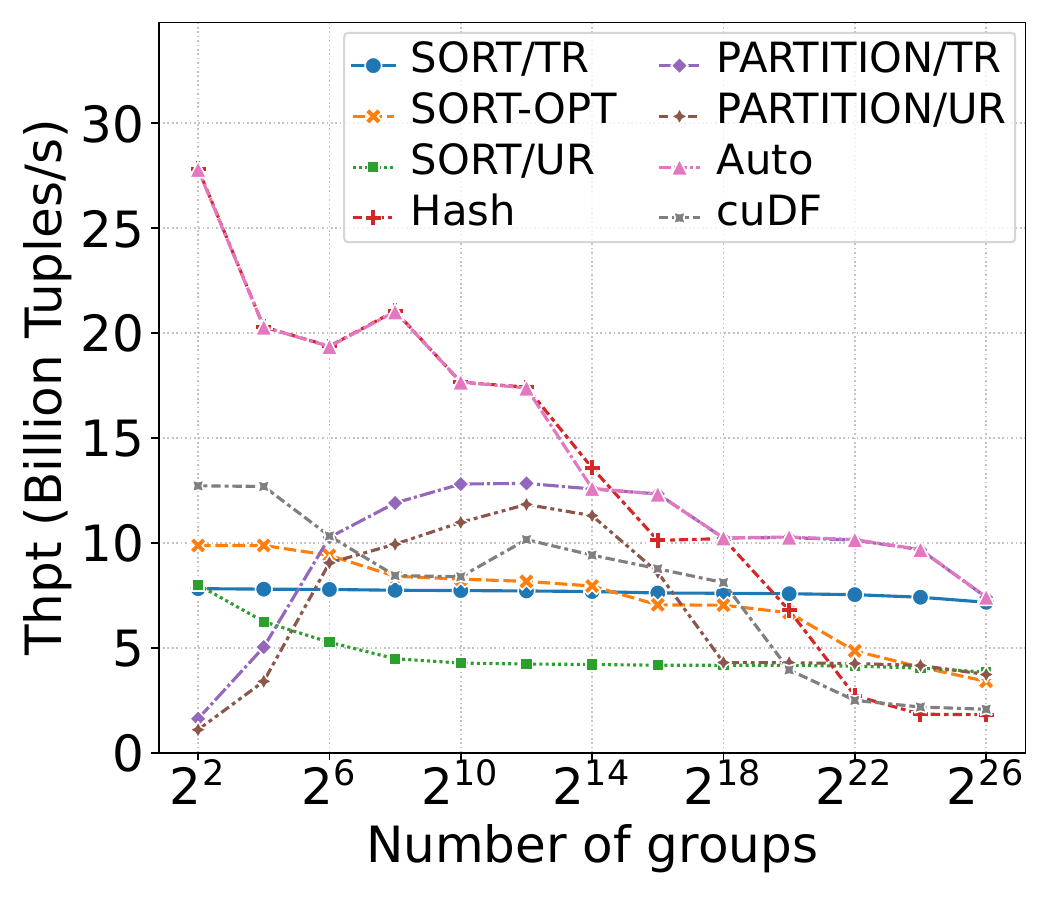}
        \caption{4-byte group key (A100).}
        \label{fig:groupby-4b-A100}
    \end{subfigure}
    \hfill
    \begin{subfigure}[b]{0.48\columnwidth}
        \centering
        \includegraphics[width=\columnwidth]{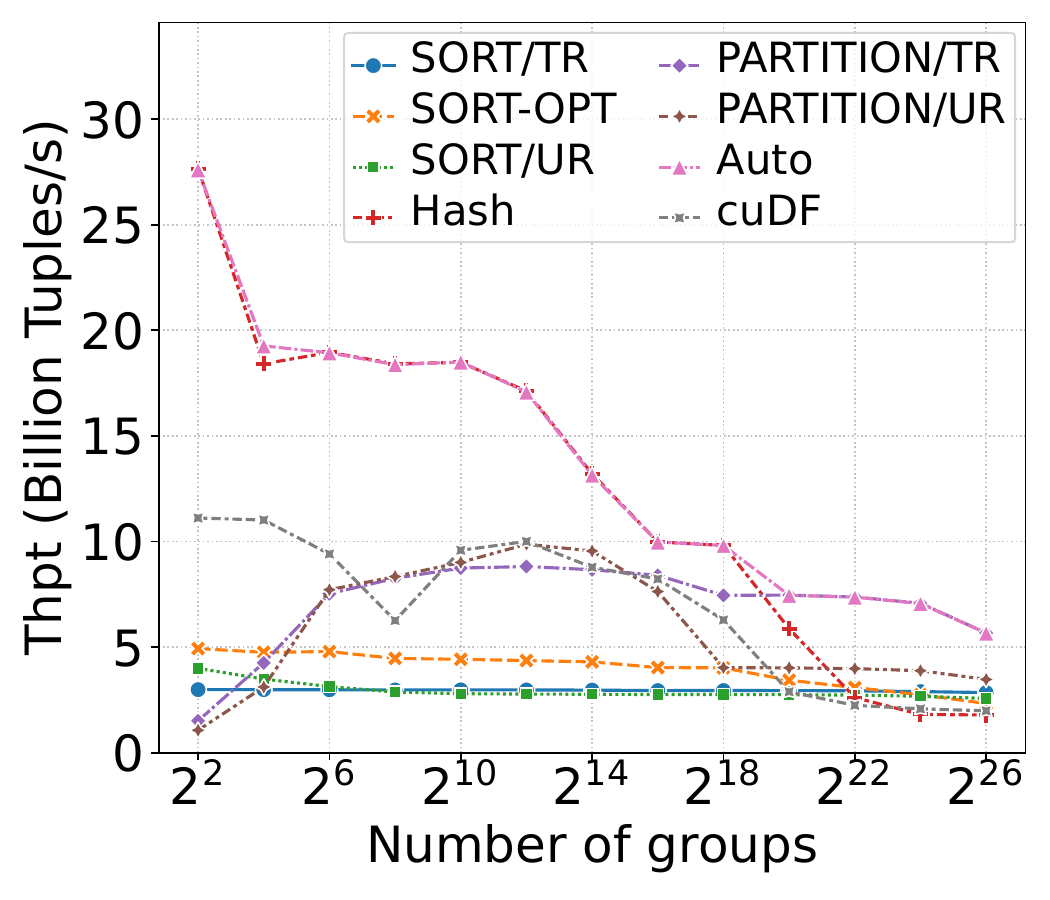}
        \caption{8-byte group key (A100).}
        \label{fig:groupby-8b-A100}
    \end{subfigure}
    \caption{Group-by microbenchmarks.}
    \label{fig:groupby}
\end{figure}

We evaluate hash-based, sort-based, and partition-based group-by implementations, including GFTR/GFUR variants and an optimized sort-based variant (SORT-OPT) that applies dictionary-encoding optimization on top of SORT/TR.
To demonstrate the effectiveness of the runtime adaptive execution for group-by (Section~\ref{sec:ndv}), we also include a series in which the algorithm is selected by calculating the HLL++ sketch. The time to compute the sketch is also included.
Unless otherwise stated, the number of rows is fixed to $2^{28}$, and the number of aggregations is fixed to 2. We denote the number of groups (i.e., group cardinality) as $g$. The results are shown in Figure~\ref{fig:groupby}.

\subsubsection{The Effect of Group Cardinality}
\label{sec:eval-groupby-card}

For 4-byte keys and payloads, hash-based group-by is the most efficient up to $g=2^{14}$ on both A100 and GH200, and in that regime it substantially outperforms cuDF, which also implements a hash-based group-by. 
Across group cardinalities, sort-based approaches are comparatively stable. GFTR outperforms GFUR for most group cardinalities (with the noted exception at $g=4$). 
SORT-OPT provides an additional advantage over vanilla SORT/TR only up to a platform-dependent threshold: up to $g=2^{14}$ on A100 and up to $g=2^{10}$ on GH200. 
PARTITION/TR performs the best when $g > 2^{20}$ on both platforms.

For 8-byte keys and payloads, the hash-based implementation remains the most efficient up to $g=2^{18}$ on A100 and up to $g=2^{14}$ on GH200; within this range, it again substantially outperforms cuDF. Sort-based approaches remain stable; however, unlike the 4-byte case, GFTR no longer consistently outperforms GFUR. SORT-OPT achieves the best overall performance and retains its advantage over vanilla SORT/UR up to $g=2^{22}$.
PARTITION/TR again performs the best when $g > 2^{20}$ on both platforms.

\subsubsection{Runtime-Adaptive Algorithm Selection}
\label{sec:groupby-auto}

The series labeled as ``Auto'' in Figure~\ref{fig:groupby} refers to the selection of the group algorithm guided by HLL++, which is part of our runtime adaptive execution mechanism introduced in Section~\ref{sec:runtime-exec}. It includes the time to compute the HLL++ sketch to estimate the number of groups and the time to execute the selected algorithm based on the estimation. The results show that our proposed runtime adaptive mechanism can almost always find the optimal algorithm. 
Moreover, it shows that HLL++ adds negligible overhead, and the estimation is accurate. 

\subsection{Expression Evaluation}
\label{sec:eval-expr}

\begin{table}[t] % Expression microbenchmarks
\centering
\caption{Expressions for microbenchmark}
\label{tab:exprs}
\resizebox{\columnwidth}{!}{%
\begin{tabular}{@{}llcl@{}}
\toprule
Id          & Expression                                               & \#Cols & Description                          \\ \midrule
\textbf{E1} & 2 * col0                                                 & 1      & Simple arithmetic                   \\
\textbf{E2} & col0 * (1 - col1)                                        & 2      & TPC-H Q1-like arithmetic expression \\
\textbf{E3} & col0 * (1 - col1) * (1 + col2)                           & 3      & TPC-H Q1-like arithmetic expression \\
\textbf{E4} & col0 * (1 - col1) - col2 * col3                          & 4      & TPC-H Q9-like arithmetic expression \\
\textbf{E5} & col0 \textgreater 134217728                              & 1      & Simple predicate                    \\
\textbf{E6} & col0 \textgreater 161061273 and col1 \textless 375809638 & 2      & Simple logical and                  \\
\textbf{E7} & \begin{tabular}[c]{@{}l@{}}(col0 == 0 and col1 == 1) \\ or (col0 == 1 and col1 == 0)\end{tabular} & 2 & Simple chained predicates \\ \bottomrule
\end{tabular}%
}
\end{table}

\begin{figure}[t] % Expression evaluation
    \centering
    \begin{subfigure}[b]{0.48\columnwidth}
        \centering
        \includegraphics[width=\columnwidth]{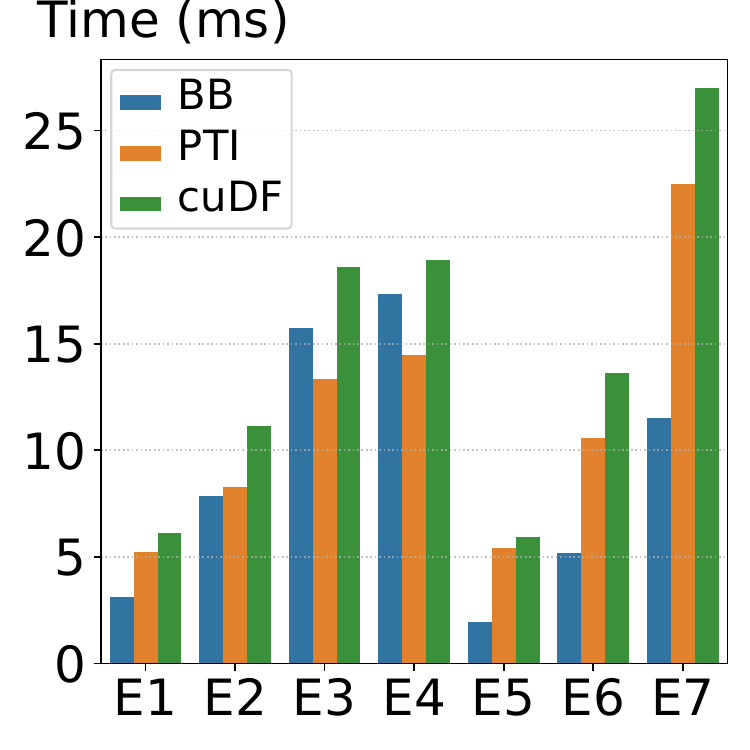}
        \caption{4-byte columns (A100)}
        \label{fig:expr-4b-A100}
    \end{subfigure}
    \hfill
    \begin{subfigure}[b]{0.48\columnwidth}
        \centering
        \includegraphics[width=\columnwidth]{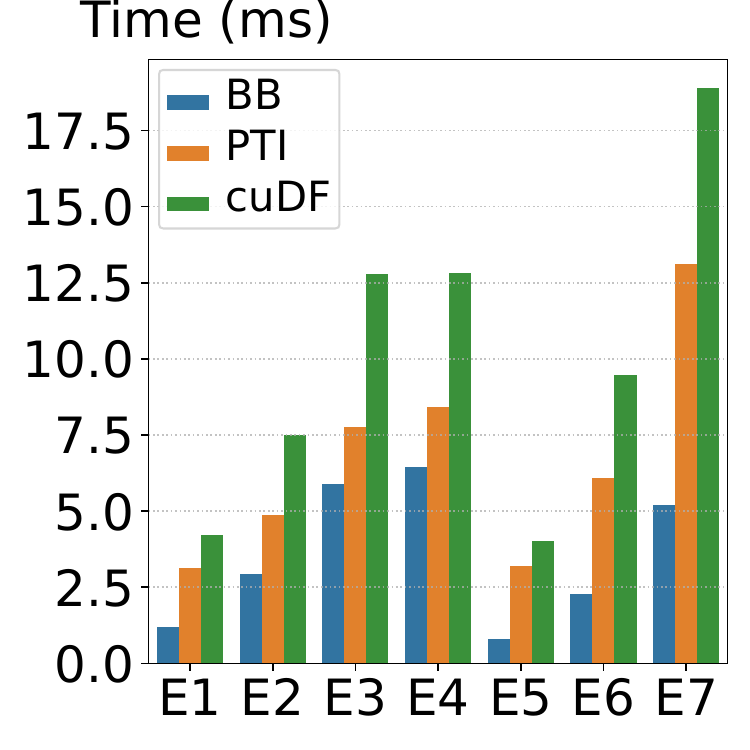}
        \caption{4-byte columns (GH200)}
        \label{fig:expr-4b-GH200}
    \end{subfigure}
    \hfill
    \begin{subfigure}[b]{0.48\columnwidth}
        \centering
        \includegraphics[width=\columnwidth]{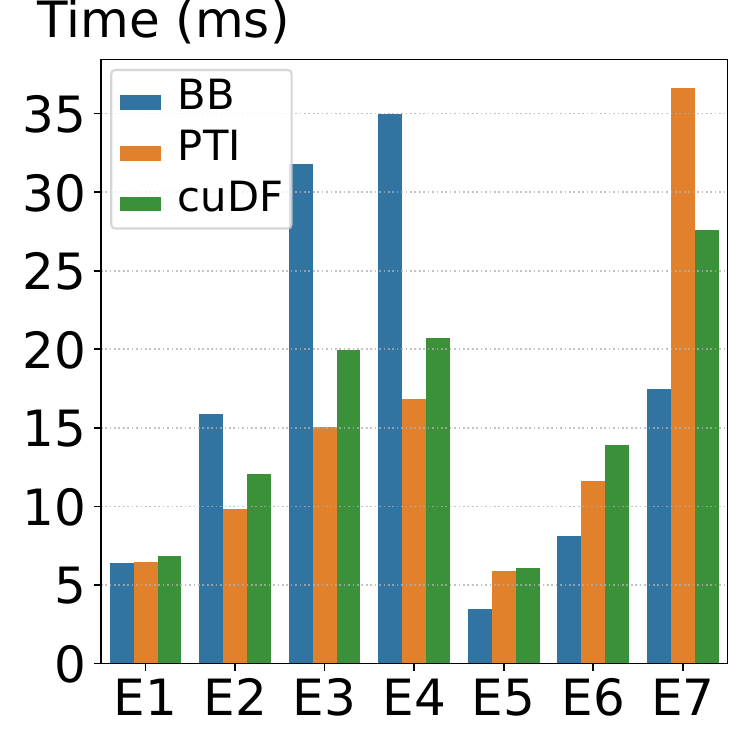}
        \caption{8-byte columns (A100)}
        \label{fig:expr-8b-A100}
    \end{subfigure}
    \hfill
    \begin{subfigure}[b]{0.48\columnwidth}
        \centering
        \includegraphics[width=\columnwidth]{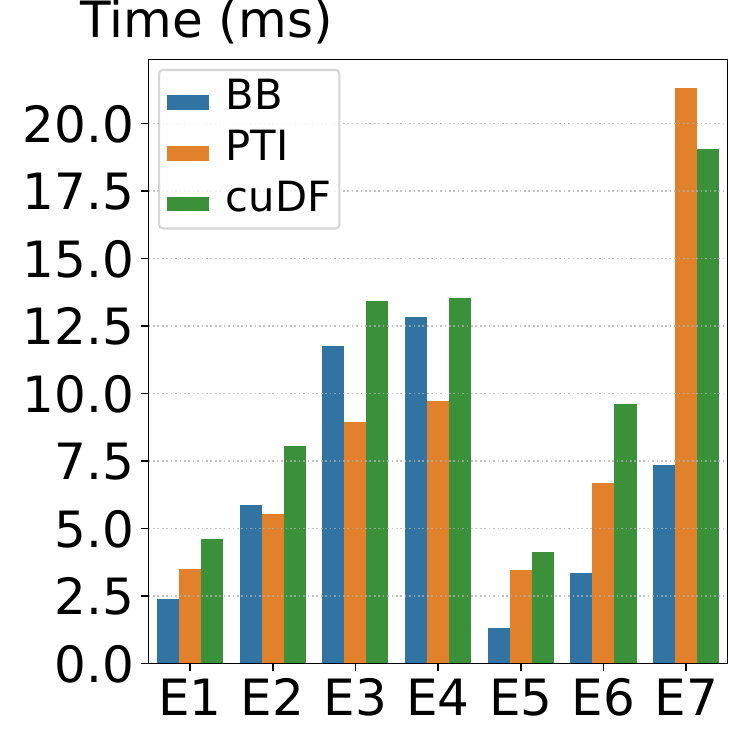}
        \caption{8-byte columns (GH200)}
        \label{fig:expr-8b-GH200}
    \end{subfigure}
    \caption{Expression evaluation microbenchmarks.}
    \label{fig:expr-mb}
\end{figure}

We compare our two expression evaluation backends (Section~\ref{sec:expr-eval}) against cuDF's AST-based expression evaluation. We do not choose the cuDF just-in-time (JIT) based evaluator because JIT incurs a high compilation overhead and is usually only good for very complex or recurring expressions. 
The expressions we use to evaluate are listed in Table~\ref{tab:exprs}, which contains a rich mix of arithmetic (E1-E4) and logical (E5-E7) operations. 
From E1 to E4 and from E5 to E7, the number of intermediate results needed to evaluate the expression increases. 
The results are shown in Figure~\ref{fig:expr-mb}. 
Figures~\ref{fig:expr-4b-A100} and Figure~\ref{fig:expr-4b-GH200} show the performance when all the columns involved are 4-byte wide. 
As the number of columns and the complexity of the expression increase, all implementations experience a longer execution time. 
Both of our implementations outperform cuDF. 
Of seven expressions, PTI and BB outperform cuDF by up to 1.4$\times$ and 3$\times$ on A100 and 1.6$\times$ and 5$\times$ on GH200. 
BB has a bigger performance advantage over PTI and cuDF on GH200. 
For columns of 8-byte width (Figure~\ref{fig:expr-8b-A100} and~\ref{fig:expr-8b-GH200}), PTI still consistently outperforms cuDF for E1-E6, but loses to cuDF for E7. 
On A100, BB only outperforms cuDF for E1 and logical expressions (E5-E7). 
As the size of intermediate results doubles for 8-byte columns when processing E1-E4, the execution time of BB also increases by around a factor of 2, while the PTI and cuDF experience a much less dramatic performance degradation. 
For E5-E7, BB is less affected because the intermediate results are boolean values, which remain the same size for both the 4- and 8-byte cases.
On GH200, due to the much higher memory bandwidth, materializing the intermediate results becomes less costly, which makes BB consistently outperform cuDF.

The relative performance between BB and PTI is very different depending on the GPUs. 
BB is more sensitive to the size of the intermediate result in A100. 
This implies that when choosing a suitable implementation, the hardware specification must also be taken into account. 
In general, BB is preferred for expressions with predominantly logical operations, where the size of intermediate results is small. 
For GPUs with a lower memory bandwidth, like A100, PTI is preferred for arithmetic-heavy expressions; for GPUs with a higher memory bandwidth, BB is preferred for low intermediate result sizes (for example, 4-byte columns). 

\begin{figure}[t] % string prefix matching
    \centering
    \begin{subfigure}[b]{0.48\columnwidth}
        \centering
        \includegraphics[width=\columnwidth]{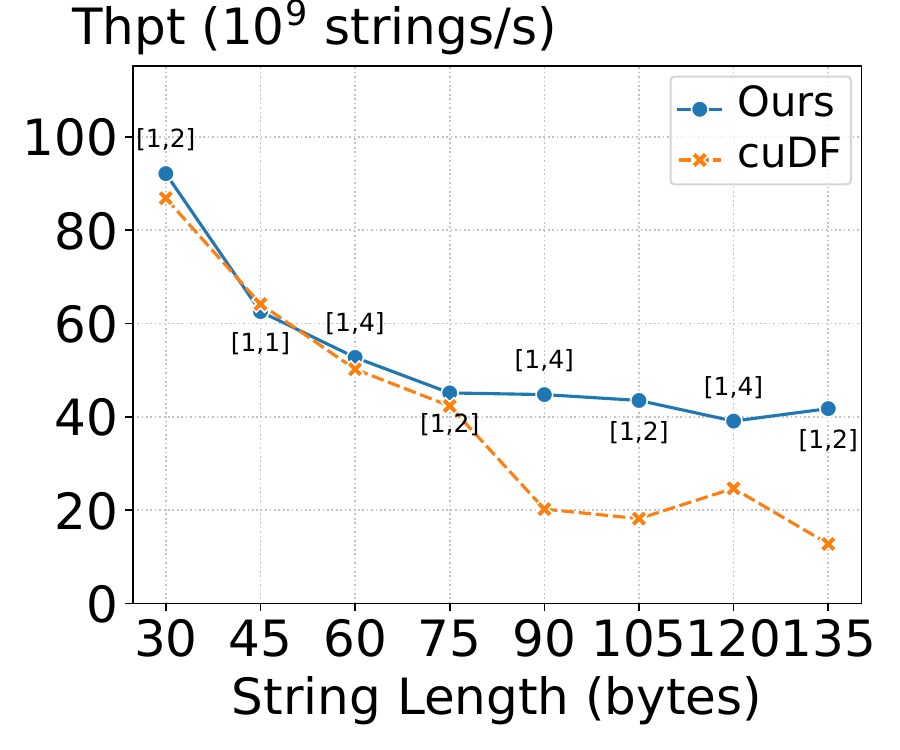}
        \caption{5-byte prefix matching (aligned)}
        \label{fig:string-startswith-5-aligned}
    \end{subfigure}
    \hfill
    \begin{subfigure}[b]{0.48\columnwidth}
        \centering
        \includegraphics[width=\columnwidth]{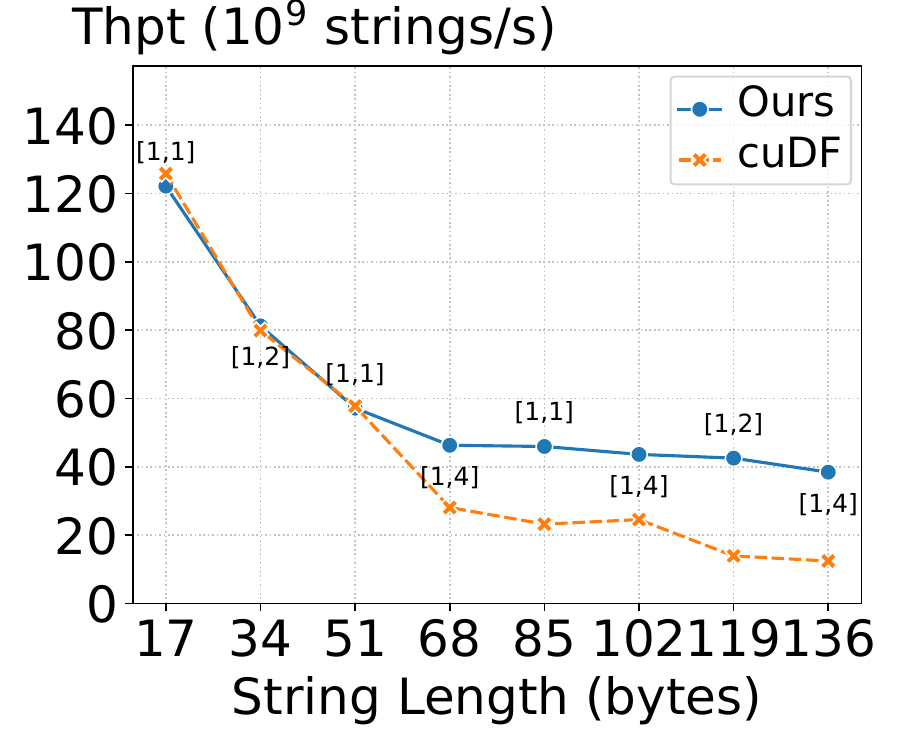}
        \caption{5-byte prefix matching (unaligned)}
        \label{fig:string-startswith-5-unaligned}
    \end{subfigure}
    \hfill
    \begin{subfigure}[b]{0.48\columnwidth}
        \centering
        \includegraphics[width=\columnwidth]{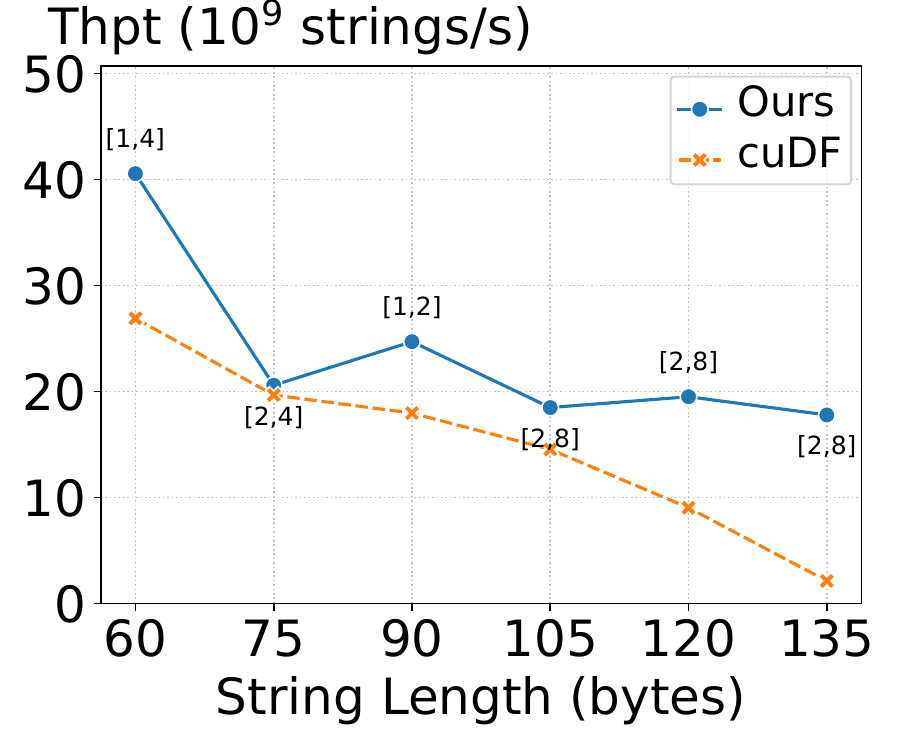}
        \caption{50-byte prefix matching (aligned)}
        \label{fig:string-startswith-50-aligned}
    \end{subfigure}
    \hfill
    \begin{subfigure}[b]{0.48\columnwidth}
        \centering
        \includegraphics[width=\columnwidth]{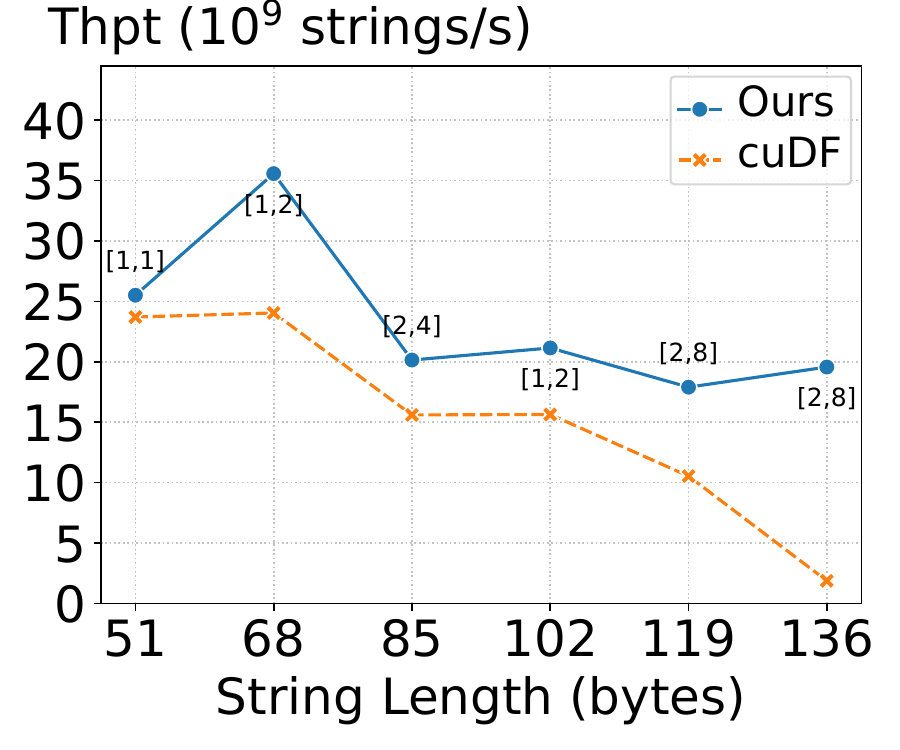}
        \caption{50-byte prefix matching (unaligned)}
        \label{fig:string-startswith-50-unaligned}
    \end{subfigure}
    \caption{String prefix matching microbenchmarks (GH200).}
    \label{fig:string-mb-GH200}
\end{figure}

\begin{figure}[t] % string exact matching
    \begin{subfigure}[b]{0.48\columnwidth}
        \centering
        \includegraphics[width=\columnwidth]{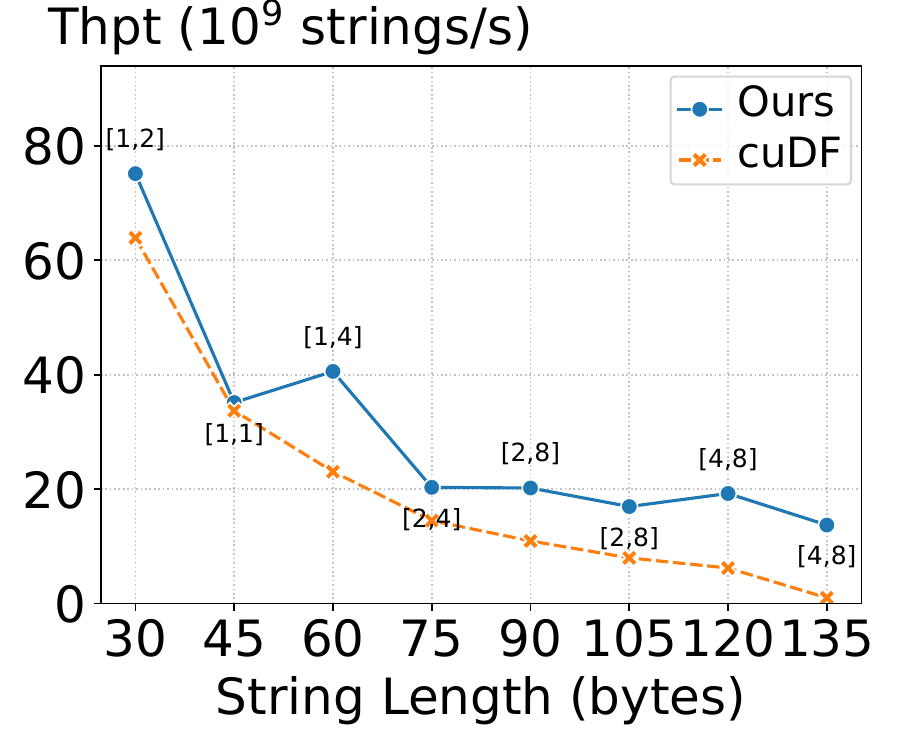}
        \caption{Full matching (aligned)}
        \label{fig:string-equal-aligned-GH200}
    \end{subfigure}
    \hfill
    \begin{subfigure}[b]{0.48\columnwidth}
        \centering
        \includegraphics[width=\columnwidth]{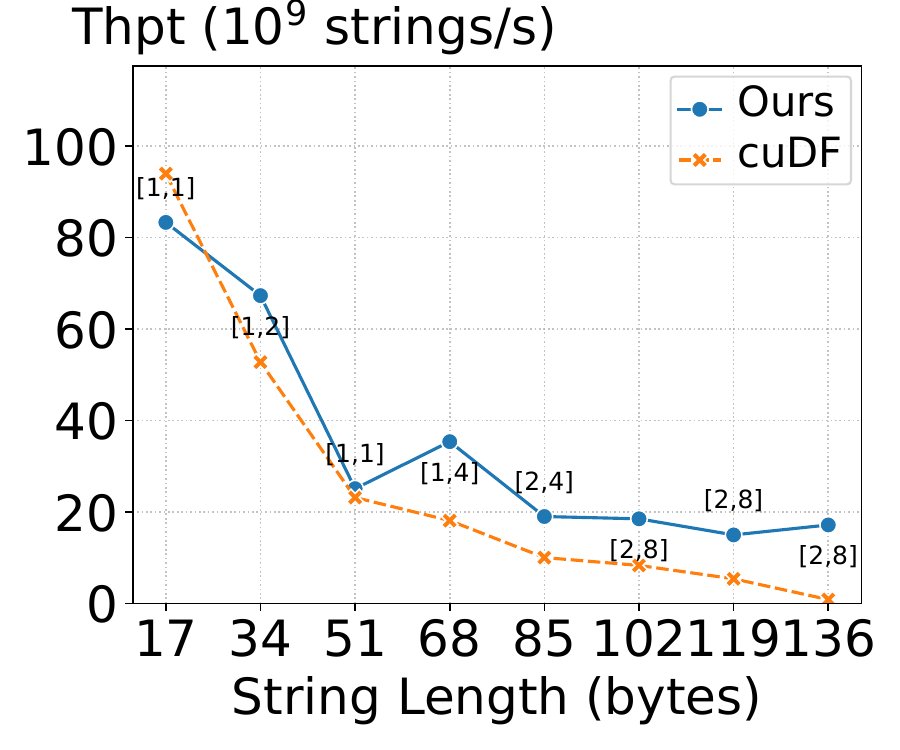}
        \caption{Full matching (unaligned)}
        \label{fig:string-equal-unaligned-GH200}
    \end{subfigure}
    \caption{String exact matching microbenchmark (GH200).}
    \label{fig:str-equal}
\end{figure}

\begin{figure}[t!] % heatmap
    \centering
    \begin{subfigure}[b]{0.48\columnwidth}
        \centering
        \includegraphics[width=\columnwidth]{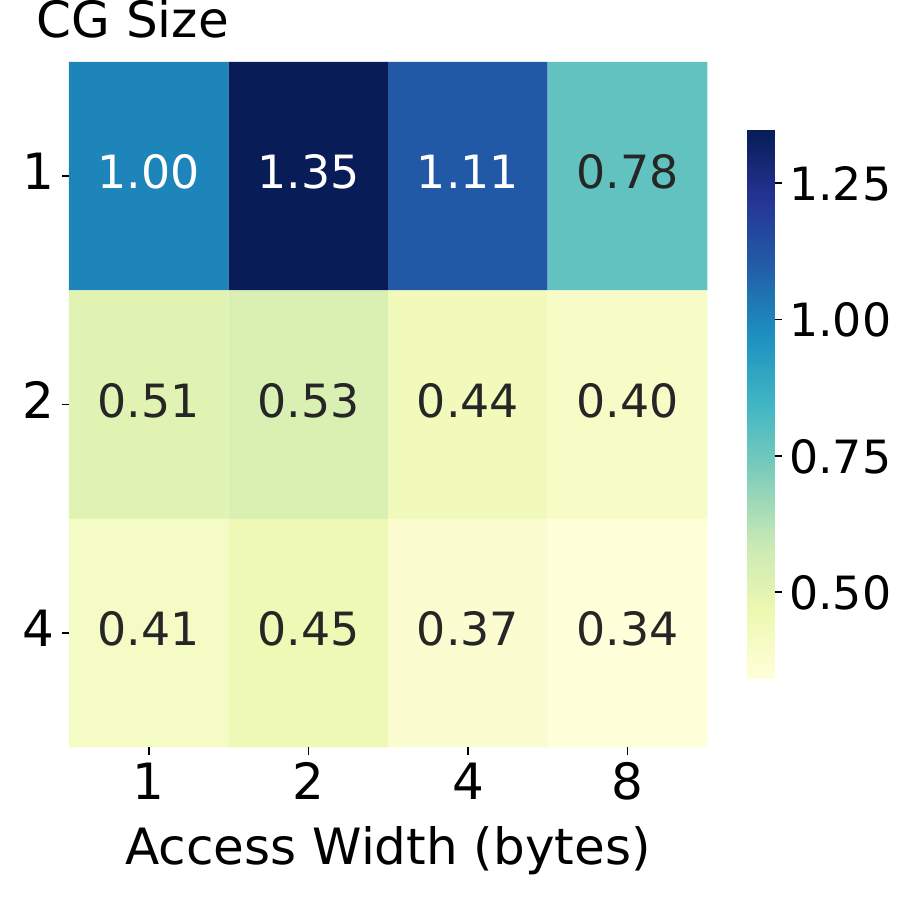}
        \caption{String length = 30}
        \label{fig:string-heatmap-30}
    \end{subfigure}
    \hfill
    \begin{subfigure}[b]{0.48\columnwidth}
        \centering
        \includegraphics[width=\columnwidth]{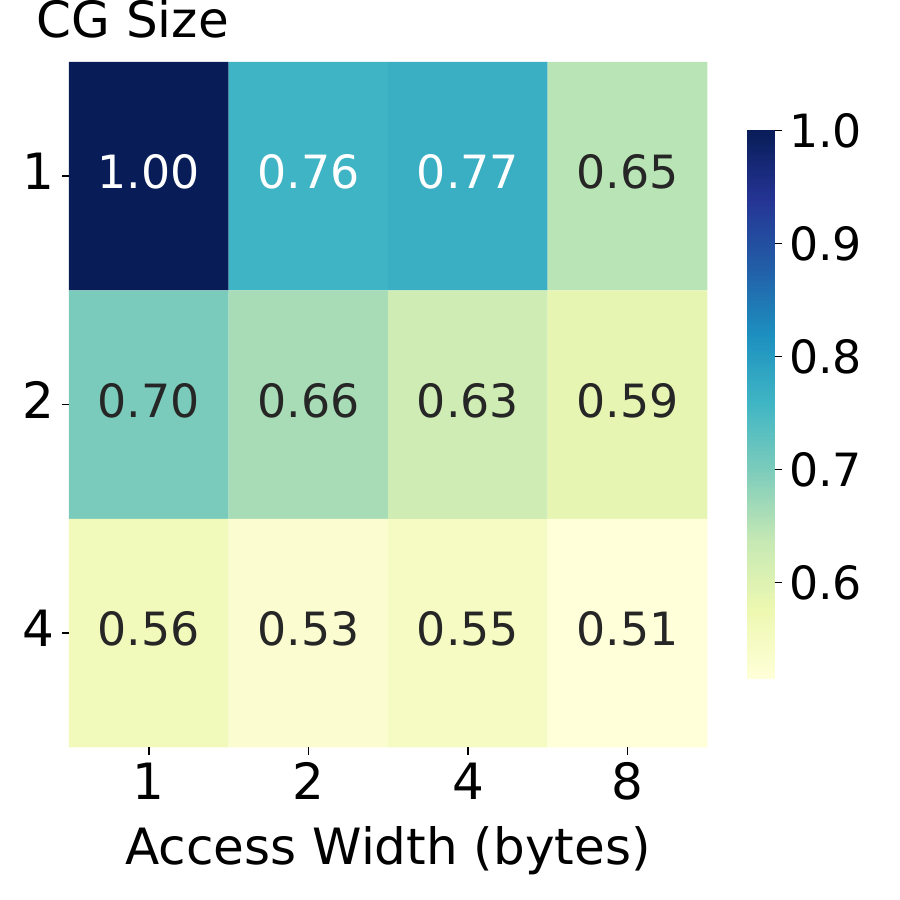}
        \caption{String length = 45}
        \label{fig:string-heatmap-45}
    \end{subfigure}
    \hfill
    \begin{subfigure}[b]{0.48\columnwidth}
        \centering
        \includegraphics[width=\columnwidth]{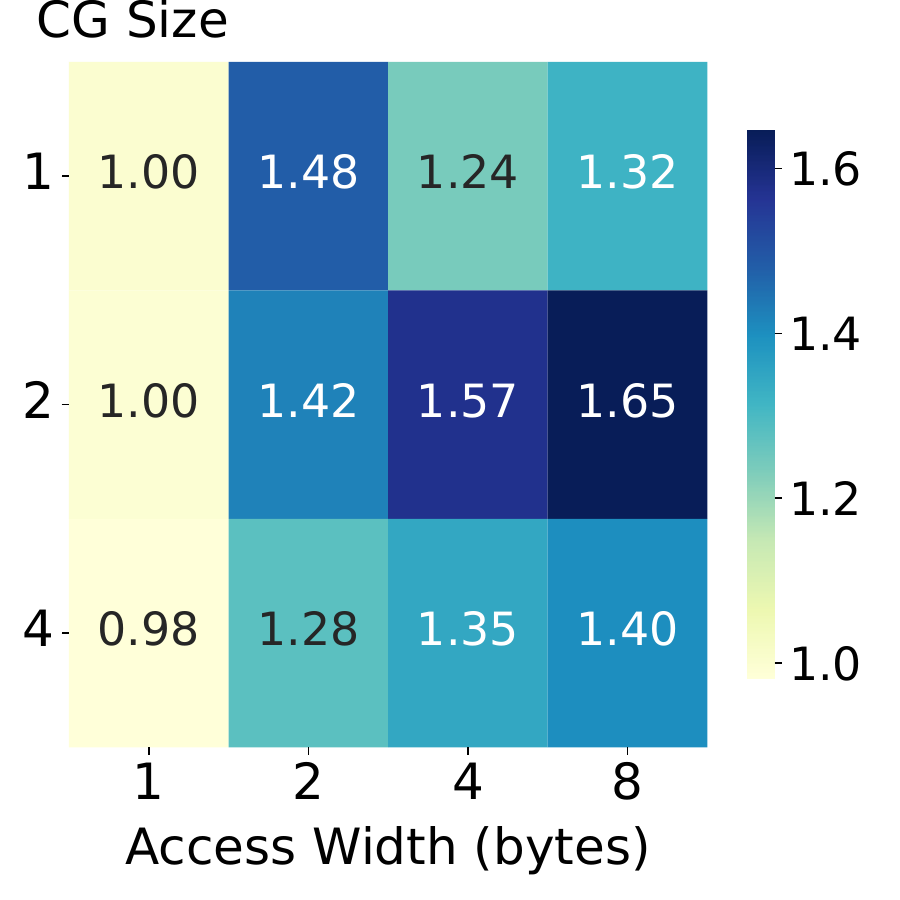}
        \caption{String length = 90}
        \label{fig:string-heatmap-90}
    \end{subfigure}
    \hfill
    \begin{subfigure}[b]{0.48\columnwidth}
        \centering
        \includegraphics[width=\columnwidth]{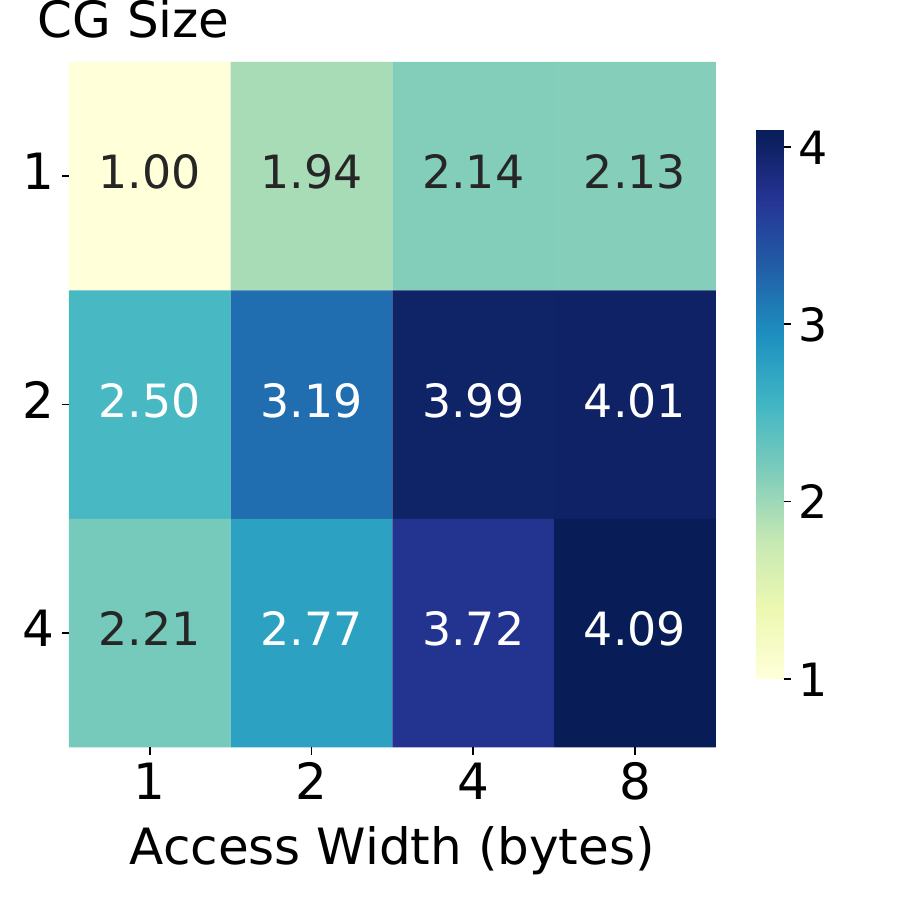}
        \caption{String length = 135}
        \label{fig:string-heatmap-135}
    \end{subfigure}
    \caption{Effect of parallelism and packed access width on full string matching (GH200).}
    \label{fig:string-heatmap}
\end{figure}

\subsection{String Processing}
\label{sec:eval-string}

We evaluate prefix/suffix, exact, and substring matching. 
If not specifically stated, we measure with $2^{27}$ strings on both platforms for a fair comparison.
Without loss of generality, when generating the column of strings, we always generate fixed-length strings for easier analysis. 
In order not to be misled by the benefit of accidental alignment, 
we also study the case with unfavorable alignments. 

\subsubsection{Prefix/Suffix/Exact Matching}
\label{sec:string-prefix-suffix-full}

In this set of experiments, we study the performance of prefix/suffix and exact string matching.
For the Arrow string format, prefix and suffix matching are equivalent; therefore, we only include the results for prefix matching.

Figures~\ref{fig:string-startswith-5-aligned}--\ref{fig:string-startswith-50-unaligned} show the results of prefix matching over two pattern lengths, 5 and 50. Figures~\ref{fig:string-startswith-5-aligned} and Figure~\ref{fig:string-startswith-50-aligned} increase the length of the string from 30/60 bytes with a step size of 15, which often creates favorable alignments.
On the other hand, Figure~\ref{fig:string-startswith-5-unaligned} and Figure~\ref{fig:string-startswith-50-unaligned} increment the string length from 17/51 bytes with a step size of 17, creating less favorable alignments.
Our implementation exposes two tuning parameters for prefix/suffix/full matching: the number of threads per string (parallelism) and the packed access width; the data points of our method in the figures are annotated with the best-performing configuration in the form $[x,y]$ where $x$ is parallelism and $y$ is access width. 
This assists in demonstrating how the best configuration changes with the change of string lengths. 

On GH200, Eiger is consistently faster than cuDF, and the advantage becomes more pronounced as the length of the string increases: Eiger’s throughput degrades more slowly with the length of the string. On A100, performance (not shown) is generally similar to cuDF across the same sweep, without a pronounced degradation trend relative to cuDF. 

For short patterns in prefix/suffix matching, using one thread per string is consistently best across both platforms. The effect of access-width follows alignment: on GH200, when the string length is a multiple of 4, 4-byte access is most efficient, otherwise 2-byte access. For the long pattern case, as the string length increases, it is beneficial to increase the parallelism as well as the access width for both favorable and unfavorable alignments.
Whether the alignment is favorable or not does not have an impact on the performance. 

For exact matching (Figure~\ref{fig:str-equal}), we make observations very similar to those of prefix/suffix matching.
To better understand the effect of parallelism (cooperative group size (CG size)) and packed access width, we created the speedup heatmaps in Figure~\ref{fig:string-heatmap}.
For both GH200 and A100, increasing parallelism and access width tends to help when the string length increases. For shorter patterns (e.g., pattern length=30), additional threads provide little benefit, while 2-byte access can help. At pattern length 45, one thread and 1-byte access is most efficient, possibly because string lengths are odd. For larger patterns, the optimal parallelism differs by GPUs: at pattern length 90, GH200 prefers 2 threads with larger access width, whereas A100 already prefers 4 threads and 8-byte access; at pattern length 135, performance correlates positively with both parallelism and access width, and this trend appears earlier on A100. 

\subsubsection{Substring Matching}
\label{sec:string-substring}

\begin{figure}[t] % substring matching
    \centering
    \begin{subfigure}[b]{0.48\columnwidth}
        \centering
        \includegraphics[width=\columnwidth]{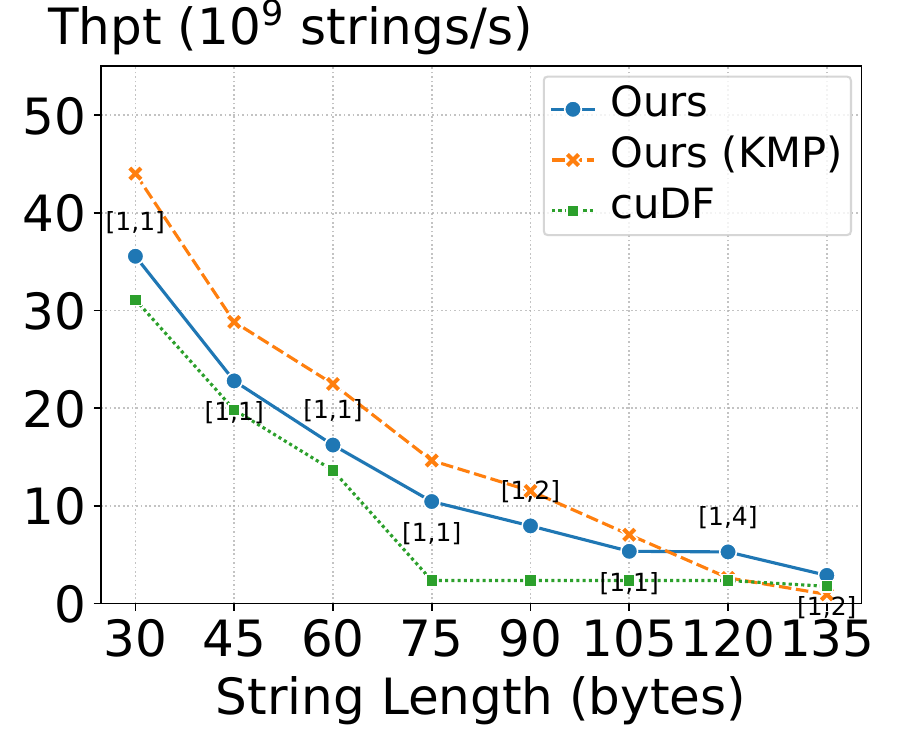}
        \caption{5-byte pattern (GH200)}
        \label{fig:string-contains-5-GH200}
    \end{subfigure}
    \hfill
    \begin{subfigure}[b]{0.48\columnwidth}
        \centering
        \includegraphics[width=\columnwidth]{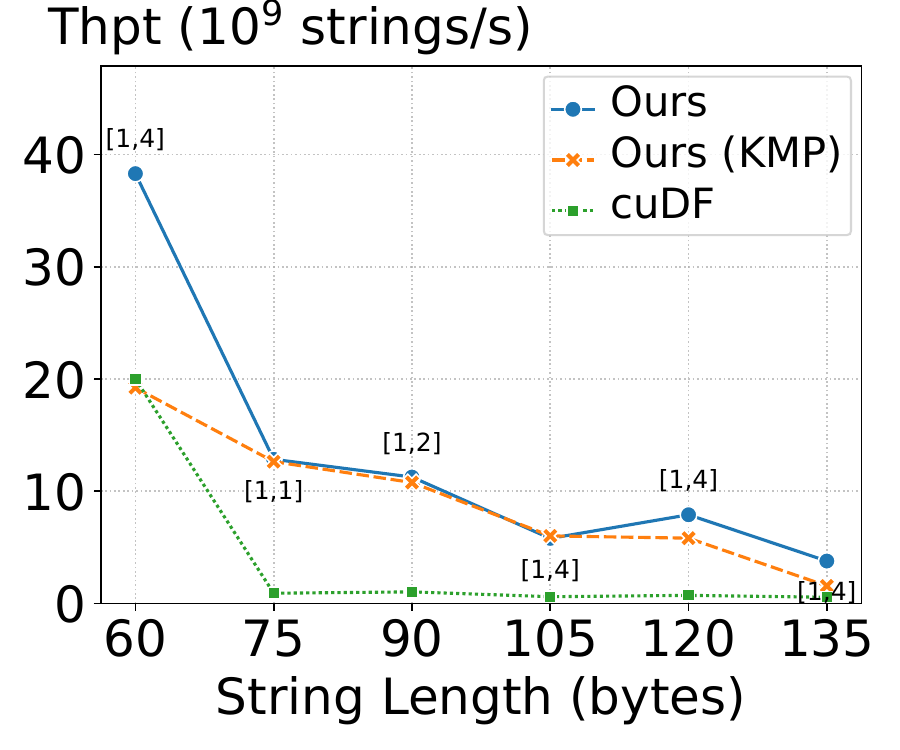}
        \caption{50-byte pattern (GH200)}
        \label{fig:string-contains-50-GH200}
    \end{subfigure}
    \hfill
    \begin{subfigure}[b]{0.48\columnwidth}
        \centering
        \includegraphics[width=\columnwidth]{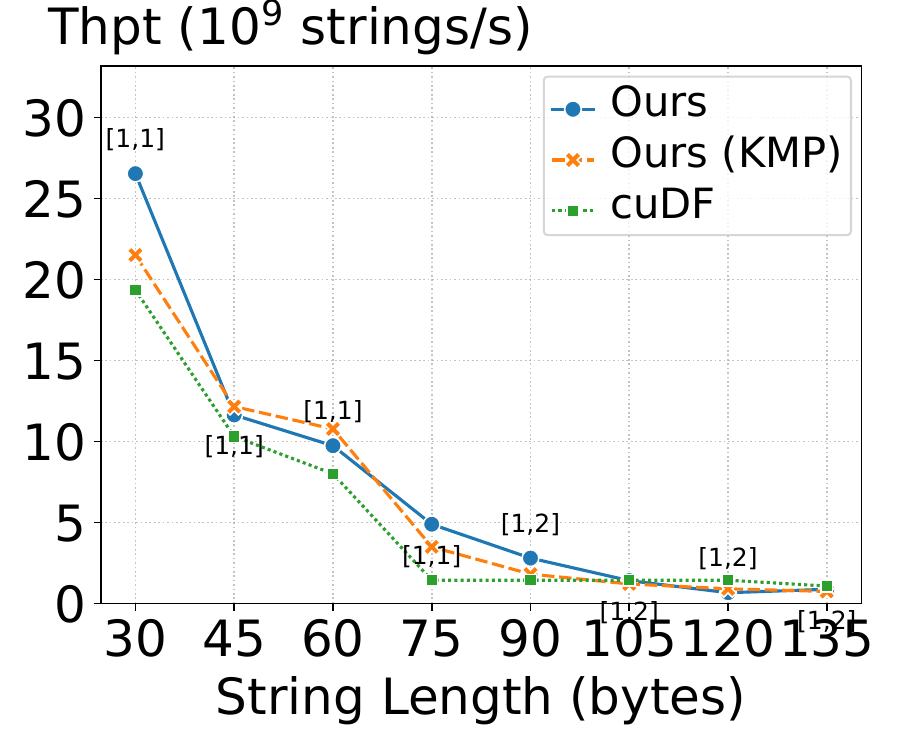}
        \caption{5-byte pattern (A100)}
        \label{fig:string-contains-5-A100}
    \end{subfigure}
    \hfill
    \begin{subfigure}[b]{0.48\columnwidth}
        \centering
        \includegraphics[width=\columnwidth]{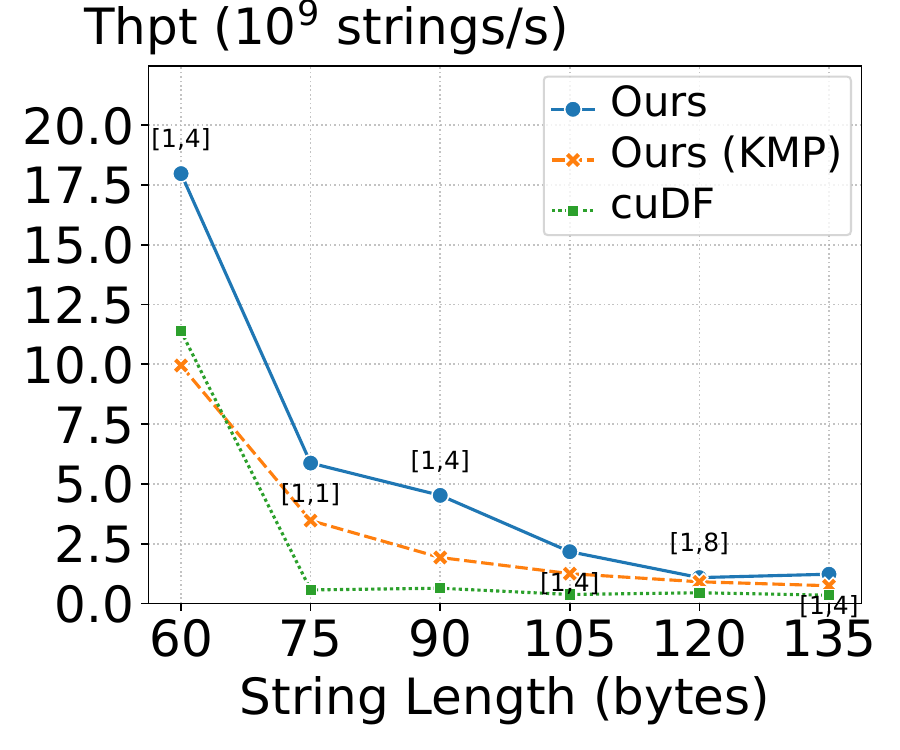}
        \caption{50-byte pattern (A100)}
        \label{fig:string-contains-50-A100}
    \end{subfigure}
    \caption{Substring matching microbenchmarks.}
    \label{fig:string-contains-A100}
\end{figure}

We evaluate substring matching with a ``vanilla'' implementation, where the packed access width is configurable, and the parallelism is fixed to 1, and a KMP-based implementation without tuning knobs. 
Figures~\ref{fig:string-contains-5-GH200}--\ref{fig:string-contains-50-GH200} and Figures~\ref{fig:string-contains-5-A100}--\ref{fig:string-contains-50-A100} show the results for GH200 and A100, respectively.

Our vanilla approach (``Ours'' in the figures) is slightly more efficient than cuDF for all string lengths and pattern lengths on both GPUs. The speedup is more prominent for larger pattern lengths. The KMP variant does not provide consistent benefits: on A100, it does not significantly help; on GH200, it slightly improves performance for string length below 120 when the pattern length is short, but when the pattern length becomes 50, it does not help and can even hurt for short strings. 

\subsubsection{String Sorting}
\label{sec:string-sort-eval}

We evaluate string sorting by varying string length, using $2^{26}$ strings. We generate each string by randomly picking each character from ``a'' to ``z''.

\begin{figure}[t] % string sorting
    \centering
    \begin{subfigure}[b]{0.48\columnwidth}
        \centering
        \includegraphics[width=\columnwidth]{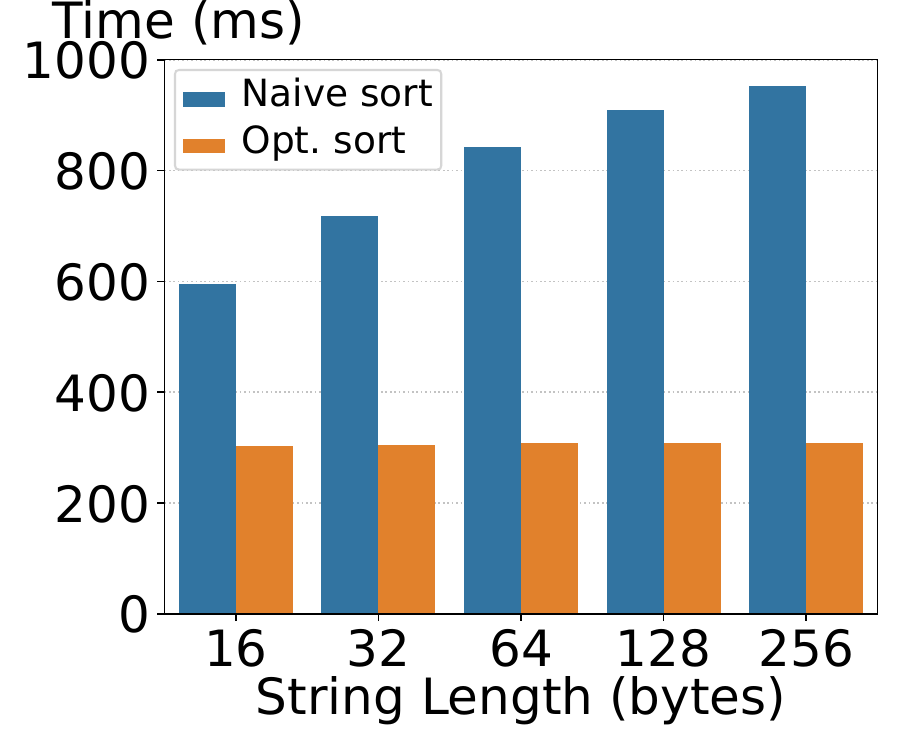}
        \caption{A100}
        \label{fig:string-sort-A100}
    \end{subfigure}
    \hfill
    \begin{subfigure}[b]{0.48\columnwidth}
        \centering
        \includegraphics[width=\columnwidth]{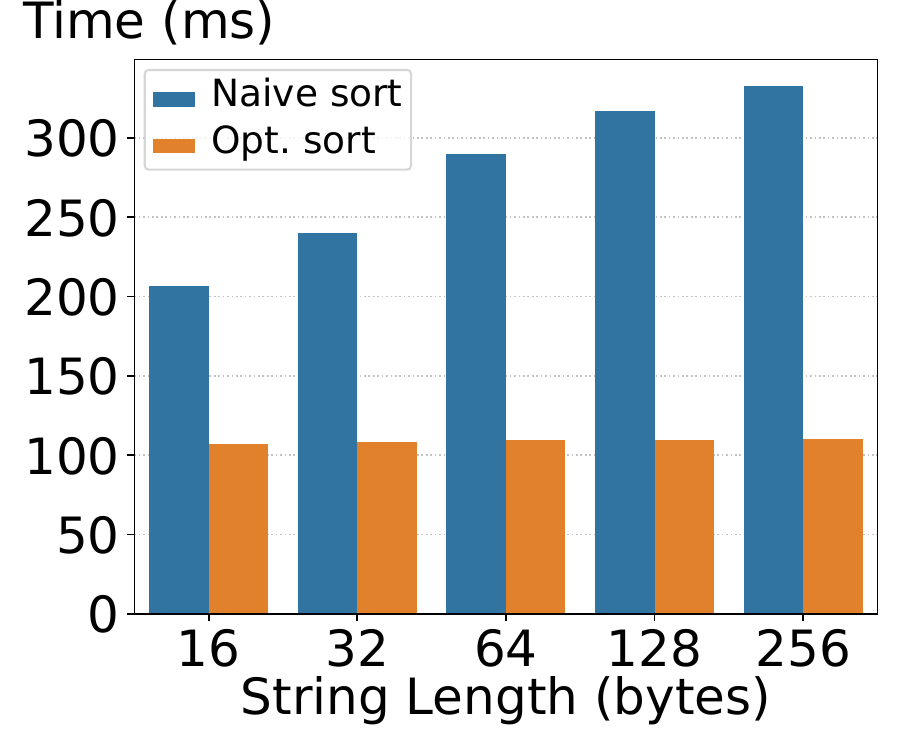}
        \caption{GH200}
        \label{fig:string-sort-GH200}
    \end{subfigure}
    \caption{String sorting microbenchmarks.}
    \label{fig:string-sort}
\end{figure}

Figure~\ref{fig:string-sort} shows that the optimized sorting using prefix extraction is more efficient than the naive merge sort on both platforms. The time of optimized sort also includes the prefix extraction time. As the string length increases, the optimized sort almost maintains the same performance, whereas the naive merge sort experiences a decrease in performance.
The performance benefit comes from extracting the prefix before the actual sorting, which allows most of the comparisons to be made based on the 4-byte prefixes only. 
This avoids reading bytes from random locations in the memory during comparison, which suffers from a worse memory access pattern. 
This is confirmed by the breakdown shown in Table~\ref{tab:stringsort-breakdown}.
The prefix extraction has a very minimal overhead but greatly reduces the execution time of the merge sort.

\begin{table}[] % string sort breakdown
\centering
\caption{Breakdown of sorting 256-byte strings (GH200).}
\label{tab:stringsort-breakdown}
\resizebox{.6\columnwidth}{!}{%
\begin{tabular}{@{}lll@{}}
\toprule
Algorithm                             & Kernels             & Time (ms) \\ \midrule
\multirow{2}{*}{Naive merge sort}     & Merge               & 261.579   \\
                                      & Block-level sort    & 69.541    \\ \midrule
\multirow{3}{*}{Optimized merge sort} & Merge               & 104.062   \\
                                      & Block-level sort    & 2.911     \\
                                      & Prefix extraction   & 1.867     \\ \bottomrule
\end{tabular}%
}
\end{table}

\begin{figure}[b] % Sort
    \centering
    \begin{subfigure}[b]{0.48\columnwidth}
        \centering
        \includegraphics[width=\columnwidth]{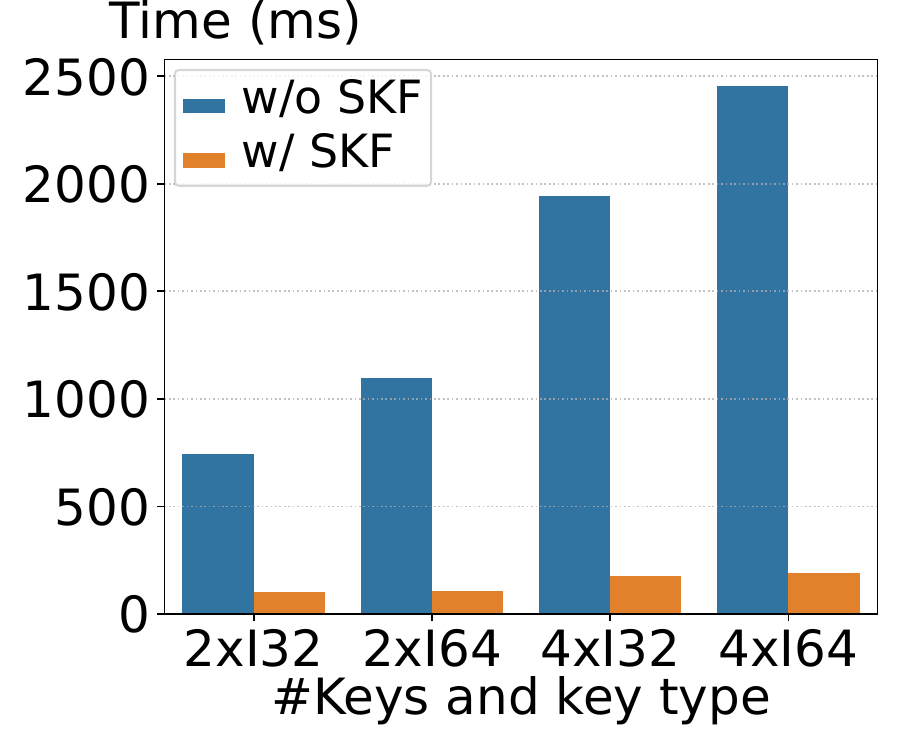}
        \caption{A100}
        \label{fig:sort-A100}
    \end{subfigure}
    \hfill
    \begin{subfigure}[b]{0.48\columnwidth}
        \centering
        \includegraphics[width=\columnwidth]{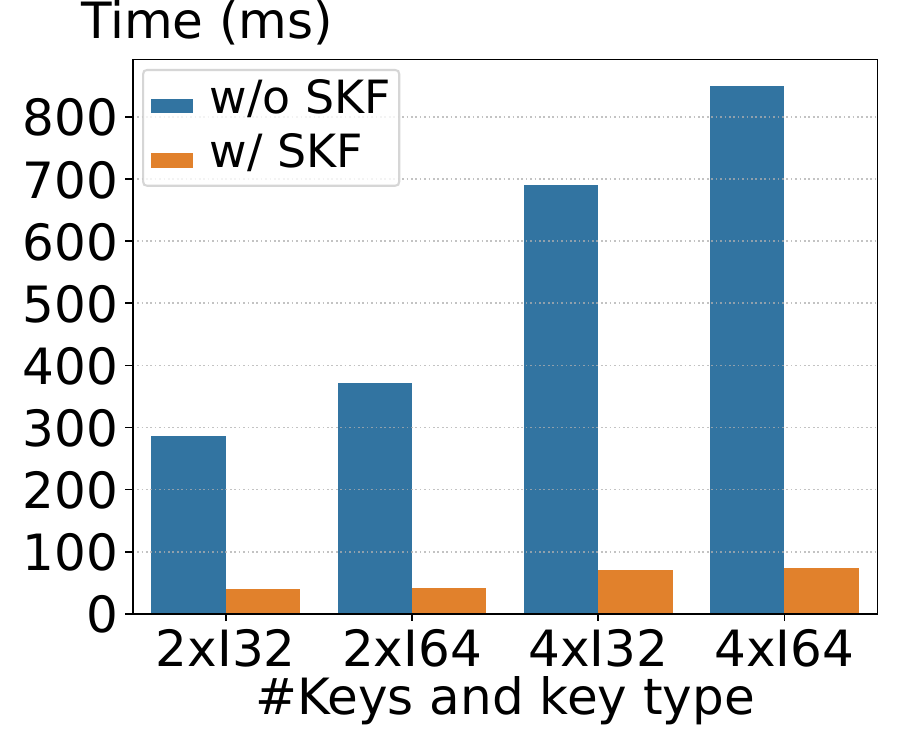}
        \caption{GH200}
        \label{fig:sort-GH200}
    \end{subfigure}
    \caption{Sort microbenchmark. SKF = smart key fusion. 2xI32 means two columns of 4-byte keys.}
    \label{fig:sort}
\end{figure}

\subsection{Sorting}
\label{sec:eval-sort}
In this experiment, we focus on the performance of multi-key sorting, commonly seen in queries, and demonstrate the effectiveness of Eiger's smart key fusion (SKF) mechanism.
We fix the number of rows to be $2^{28}$ and vary the number of key columns and the type of keys (I32 and I64).
Half of the key columns can be compressed because of the narrow value distribution (i.e., $\max - \min$ is small), while the other half has a sparse distribution but a low number of distinct values, where HyperLogLog++ can identify this case, and dictionary encoding will be used to compress the column. 

Figure~\ref{fig:sort} shows the results of this experiment. The time reported for SKF also includes the time spent collecting statistics and fusing the keys. The implementation without SKF is the merge sort. 
The results show that SKF significantly improves performance by up to 13$\times$ on A100 and 12$\times$ on GH200 compared to the merge sort approach, even with the overhead of statistics computation and key fusion.
The performance gain comes from the fact that fusing keys together enables us to use the more efficient radix sort, which has a lower time complexity than the merge sort. Fusing the keys together also results in fewer bytes being read, in contrast to the merge sort, where each comparison needs to read multiple keys. We detail the time breakdown in Table~\ref{tab:sort-breakdown}. It is obvious that the radix sort is more than an order of magnitude more efficient than the merge sort in this case, and the overhead of calculating statistics and compressing keys is well paid off. 

\begin{table}[t] % Sort breakdown
\centering
\caption{Breakdown of sorting tables (4xI32) (GH200).}
\label{tab:sort-breakdown}
\resizebox{.6\columnwidth}{!}{%
\begin{tabular}{@{}lll@{}}
\toprule
Algorithm                         & Kernels             & Time (ms) \\ \midrule
\multirow{2}{*}{Merge sort}       & Merge               & 612.311    \\
                                  & Block-level sort    & 35.791    \\ \midrule
\multirow{4}{*}{SKF + Radix sort} & Radix sort          & 12.224    \\
                                  & Dictionary encoding & 6.936     \\
                                  & Fusing keys         & 3.393     \\
                                  & Computing HLL++     & 1.061     \\ \bottomrule
\end{tabular}%
}
\vspace{-1em}
\end{table}

\subsection{TPC-H Benchmark}
\label{sec:eval-tpch}
\begin{figure*}[t]
    \centering
    \includegraphics[width=\textwidth]{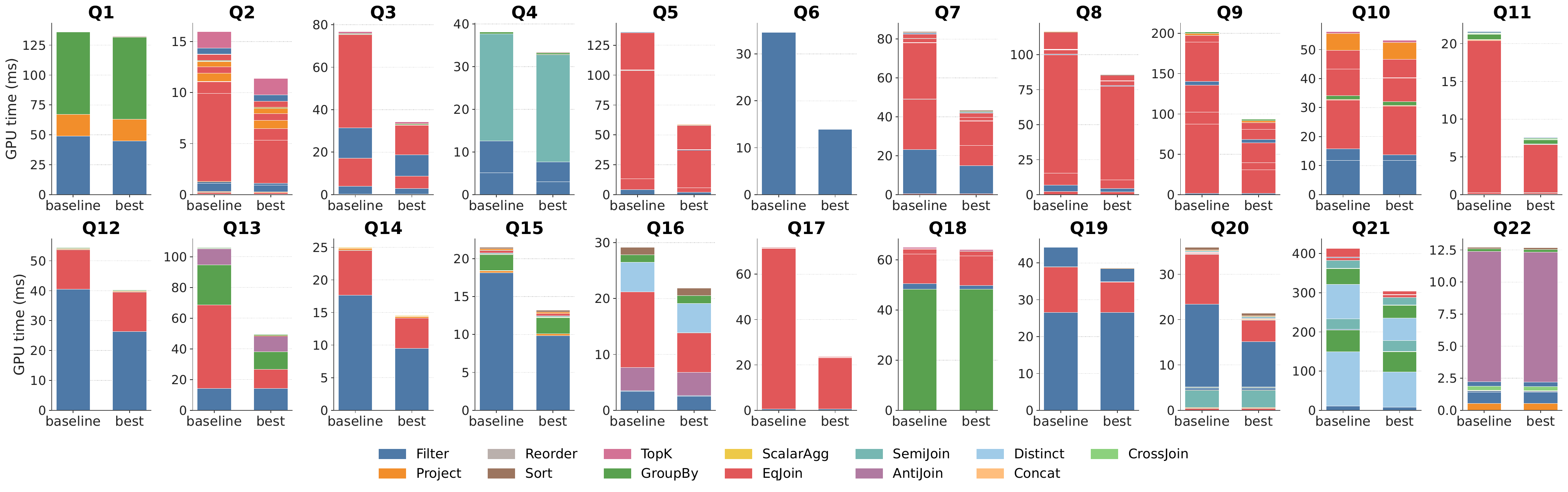}
    \caption{TPC-H SF=100 breakdown (GH200).}
    \label{fig:tpch-breakdown}
\end{figure*}

In this section, we evaluate Eiger using the standard TPC-H query benchmark~\cite{tpch}.
Table~\ref{tab:tpch-eval} shows the execution time per-query at three different scale factors, as well as the comparison with cuDF.
For the cuDF baseline, we use Maximus~\cite{kabic25-maximus}, an open-source query execution engine that integrates cuDF for GPU execution.
Data movement times between the CPU and GPU are excluded, since the evaluation aims at comparing the efficiency of GPU execution.
As discussed in Section~\ref{sec:eiger-overview}, Eiger provides different implementations for almost all operators to adapt to a wide range of workloads. Therefore, we include two different measurements from Eiger, baseline and best. Eiger (baseline) uses the same algorithm and configuration for all occurrences of the same operation. Specifically, hash-based algorithms are used for join and grouped aggregation, PTI-based expression evaluation is used for selection and projection, and smart-key-fusion (SKF) optimization is disabled for sorting and grouped aggregation. The selection of baseline implementations mirrors the cuDF algorithms. On the other hand, Eiger (best) combines the best-performing implementation of each operator. For brevity, we only show Eiger (best) vs. cuDF in Table~\ref{tab:tpch-eval}, and Eiger (best) vs. Eiger (baseline) is shown in Figure~\ref{fig:tpch-breakdown}.

\begin{table}[t]
\caption{TPC-H evaluation on GH200. Time is in ms.}
\label{tab:tpch-eval}
\resizebox{\columnwidth}{!}{%
\begin{tabular}{@{}lrrr|rrr|rrr@{}}
\toprule
    & \multicolumn{3}{c|}{SF=10}                                                & \multicolumn{3}{c|}{SF=30}                                                & \multicolumn{3}{c}{SF=100}                                                \\ \cmidrule(l){2-10} 
    & \multicolumn{1}{l}{Eiger (best)} & \multicolumn{1}{l}{cuDF} & Speedup     & \multicolumn{1}{l}{Eiger (best)} & \multicolumn{1}{l}{cuDF} & Speedup     & \multicolumn{1}{l}{Eiger (best)} & \multicolumn{1}{l}{cuDF} & Speedup     \\ \cmidrule(l){2-10} 
Q1  & 14.36                            & 24.01                    & 1.7$\times$                  & 40.54                            & 67.59                    & 1.7$\times$                  & 132.22                           & 219.51                   & 1.7$\times$                 \\
Q2  & 5.66                             & 6.73                     & 1.2$\times$                  & 7.96                             & 8.86                     & 1.1$\times$                  & 11.4                             & 15.89                    & 1.4$\times$                 \\
Q3  & 6.11                             & 11.11                    & 1.8$\times$                  & 13.14                            & 23.66                    & 1.8$\times$                  & 34.2                             & 74.75                    & 2.2$\times$                 \\
Q4  & 3.91                             & 4.67                     & 1.2$\times$                  & 10.55                            & 10.81                    & 1$\times$                    & 33.38                            & 31.83                    & 1$\times$                   \\
Q5  & 8.28                             & 13.45                    & 1.6$\times$                  & 19.77                            & 34.49                    & 1.7$\times$                  & 58.94                            & 110.44                   & 1.9$\times$                 \\
Q6  & 1.6                              & 4.42                     & 2.8$\times$                  & 4.41                             & 11.29                    & 2.6$\times$                  & 14.08                            & 34.59                    & 2.5$\times$                 \\
Q7  & 8.57                             & 14.52                    & 1.7$\times$                  & 16.69                            & 24.13                    & 1.4$\times$                  & 43.31                            & 73.45                    & 1.7$\times$                 \\
Q8  & 7.15                             & 12.78                    & 1.8$\times$                  & 15.94                            & 31.27                    & 2$\times$                    & 85.69                            & 98.62                    & 1.2$\times$                 \\
Q9  & 13.94                            & 17.95                    & 1.3$\times$                  & 34.15                            & 50.08                    & 1.5$\times$                  & 94.09                            & 165.71                   & 1.8$\times$                 \\
Q10 & 7.19                             & 9.61                     & 1.3$\times$                  & 16.13                            & 20.63                    & 1.3$\times$                  & 53.18                            & 60.69                    & 1.1$\times$                 \\
Q11 & 1.81                             & 3.36                     & 1.9$\times$                  & 3.52                             & 5.91                     & 1.7$\times$                  & 7.58                             & 14.99                    & 2$\times$                   \\
Q12 & 4.73                             & 8.81                     & 1.9$\times$                  & 12.16                            & 21.36                    & 1.8$\times$                  & 40.41                            & 67.72                    & 1.7$\times$                 \\
Q13 & 6.52                             & 23.02                    & 3.5$\times$                  & 16.82                            & 68.32                    & 4.1$\times$                  & 49.59                            & 224.66                   & 4.5$\times$                 \\
Q14 & 1.86                             & 4.4                      & 2.4$\times$                  & 4.85                             & 8.68                     & 1.8$\times$                  & 14.58                            & 24.3                     & 1.7$\times$                 \\
Q15 & 2.11                             & 4                        & 1.9$\times$                  & 4.65                             & 7.22                     & 1.6$\times$                  & 13.2                             & 17.64                    & 1.3$\times$                 \\
Q16 & 5.23                             & 17.46                    & 3.3$\times$                  & 9.53                             & 45.88                    & 4.8$\times$                  & 21.85                            & 133.31                   & 6.1$\times$                 \\
Q17 & 4.18                             & 9.49                     & 2.3$\times$                  & 8.66                             & 23.66                    & 2.7$\times$                  & 23.75                            & 73.7                     & 3.1$\times$                 \\
Q18 & 7.59                             & 8.37                     & 1.1$\times$                  & 19.68                            & 20.77                    & 1.1$\times$                  & 64.24                            & 64.46                    & 1$\times$                   \\
Q19 & 4.56                             & 7.76                     & 1.7$\times$                  & 12.39                            & 19.27                    & 1.6$\times$                  & 38.58                            & 59.86                    & 1.6$\times$                 \\
Q20 & 4.52                             & 6.73                     & 1.5$\times$                  & 9.27                             & 11.73                    & 1.3$\times$                  & 21.45                            & 26.63                    & 1.2$\times$                 \\
Q21 & 35.36                            & 52.29                    & 1.5$\times$                  & 95.07                            & 146.75                   & 1.5$\times$                  & 307.4                            & 483.89                   & 1.6$\times$                 \\
Q22 & 1.95                             & 5.25                     & 2.7$\times$                  & 4.43                             & 8.84                     & 2$\times$                    & 12.66                            & 22.26                    & 1.8$\times$                 \\ \bottomrule
\end{tabular}%
}
\end{table}

For the total runtime of 22 queries, Eiger (best) is 1.7$\times$ (SF=10), 1.8$\times$ (SF=30) and 1.8$\times$ (SF=100) faster than cuDF.
The most significant speedup is observed for Q16, Q13, and Q17. For Q16, cuDF uses the expensive merge sort for the ``count(distinct)'' aggregation while Eiger uses a more efficient hash-based approach. For Q13 and Q17, Eiger wins by using the the partition hash join.
Compared to Eiger (baseline), Eiger (best) is 1.2$\times$ (SF=10), 1.4$\times$ (SF=30), and 1.5$\times$ (SF=100) faster. This shows the importance of having multiple implementations per operator, especially for larger datasets.
To further demonstrate that the choice of operator implementations can greatly influence performance, we plot the operator time breakdown for SF=100 in Figure~\ref{fig:tpch-breakdown}. The results reveal three main influential factors that contribute to the performance gain.  
(1) Some queries (e.g., Q4, Q6, Q7, Q12, Q14, Q15, Q20) benefit from a better filter performance using the batch-based expression evaluation. The reason is that the predicates in TPC-H are commonly simple, and the intermediate results written back to the memory are only 1-byte booleans. This makes BB more efficient than PTI in evaluating expressions by reducing the interpretation overhead while not significantly increasing the memory load and store.
(2) Another group of queries (e.g., Q13, Q21) benefit from a more efficient group-by and distinct implementation. Q13 benefits from using the partition-based group-by with GFTR materialization. The group-by in the subquery has a high group cardinality, equal to the total number of rows in the customer table, making partition-based group-by the best implementation (see Section~\ref{sec:eval-groupby}). Q21's distinct operators get more efficient from the smart-key-fusion (SKF) that fuses the l\_orderkey and l\_suppkey and using a sort-based implementation. 
(3) A third group of queries benefit substantially from more efficient joins, including Q2, Q3, Q5, Q7, Q8, Q9, Q11, Q13, Q16, Q17, and Q20. Joins in these queries gain more efficiency by using the sort-merge join or partition hash join with GFUR or GFTR materialization strategies. This justifies Eiger's design principles as well as proposed optimization techniques.

Together with the microbenchmark results, we demonstrate that Eiger offers better performance due to its more efficient operator implementation, richer choice of algorithms, and the runtime adaptive execution mechanism.

\section{Related work}
\noindent\textbf{GPU-accelerated query engines.} A multitude of research prototypes and industrial systems~\cite{raydb,Hu22-tcudb,crystal_shanbhag_2020,surakav_he_2022,hong24-themis,yogatama_rethinking_2025,Yogatama25-lancelot,chrysogelos19-hetexchange,liu2022-ghive, mohr23-boss, kabic25-maximus,velox-cudf,zhang25-tqex,yuan25-vortex,Yogatama22-mordred,wu2025terabytescaleanalyticsblinkeye,Li25-cpugpudb,gqe,ozawa2026datapathfusiongpu,mageirakos2026gpugpuvectorsearch,luo2026-pystachio,presto-gpu} have explored query processing on GPUs from many different angles. 
Eiger distinguishes itself from this body of work by focusing on the operators themselves: it offers multiple implementation variants per operator, adapts the choice and configuration to the workload at runtime, and can be integrated into these engines to further improve their performance. Among these systems, Themis~\cite{hong24-themis} is also adaptive, but it targets load imbalance across threads and warps within an operator, whereas Eiger adapts the choice of algorithms and configurations to the data being processed.

\noindent\textbf{Studies of GPU operators.} 
Many studies have focused on performance analysis~\cite{Cao23-gpudb} and optimizing various database operations on a single GPU, including join~\cite{wu25-gpu-joins-groupby,sioulas19-partitioned-radix-join,Kaldewey12-gpujoin,sun23-mmjoin,Lutz22-tritonjoin,Rui17-fastequijoin,Paul20-revisit-gpujoin,Lutz20-nonpartitionedjoin,hong24-themis,Doraiswamy23-graphicsjoin}, group-by~\cite{wu25-gpu-joins-groupby,Diego18-groupby,Tomas15-groupby,rosenfeld-hash-groupby,kroviakov-crossdevice}, string processing~\cite{Sitaridi16_gpudb_string}, encoding and decoding~\cite{hepkema25-galp,boeschen24-golap,huang2025gpuaccelerationsqlanalytics,Shanbhag22-compression}, etc. 
Wu et al.~\cite{wu25-gpu-joins-groupby} and Sioulas et al.~\cite{sioulas19-partitioned-radix-join} propose two variants of partitioned hash join. 
The former is implemented in Eiger because of its compatibility with the GFTR technique. 
Wu et al. in the same work also propose three families of group-by implementations, including hash-based, sort-based, and partition-based. Eiger incorporates and further improves them and introduces the use of the HyperLogLog++ sketch to guide the algorithm selection. 
Shanbhag et al.~\cite{crystal_shanbhag_2020} propose a block-based processing paradigm and implement basic filtering, projection, join, and group-by operations. However, their implementation is not generic enough to handle arbitrary expressions or input. 
Sitaridi et al.~\cite{Sitaridi16_gpudb_string} propose multiple techniques for substring matching and alternative string formats. 
Eiger improves string processing performance by leveraging capabilities of modern GPUs and the CUDA programming model, such as packed accesses and cooperative groups.

\noindent\textbf{Adaptive query processing.} Runtime adaptivity has a long tradition in CPU databases, ranging from Eddies that reroute individual tuples among operators during execution~\cite{avnur00-eddies} to mid-query re-optimization and a broad spectrum of other adaptive query processing techniques~\cite{deshpande07-adaptive-qp}. This line of work adapts at the level of the query plan, reordering operators, switching plans, or deferring plan choices, and is therefore naturally situated inside a database system. Eiger, as a library, adapts \emph{within} operators instead: it chooses the implementation, configuration, and data representation of each individual operator. The signals also differ: while prior CPU techniques mostly react to information observed as a byproduct of execution, such as cardinalities and selectivities, Eiger proactively computes dedicated statistics over intermediate data (min/max/mean and HyperLogLog++ sketches) and even compresses the data on the fly, both of which are affordable because GPUs perform these computations at close to memory bandwidth.

Compared to existing work, Eiger offers multiple implementation variants for the same operator and features a runtime adaptive execution mechanism that selects among them based on lightweight data statistics. Instead of focusing on the few commonly discussed operators, such as joins, this work also studies in-depth the implementation of often overlooked but expensive operations, such as expression evaluation, string processing, and multi-key sorting.
Furthermore, we present a more comprehensive performance analysis than previous work, covering a wider range of operators and workloads and characterizing when each implementation variant wins, which helps future query optimizers develop accurate cost models.

\section{Conclusion}
In this work, we present Eiger, a high-performance library for GPU-based
data analytics built around \emph{runtime workload adaptivity}. Eiger
realizes this idea through two complementary design principles: it provides
multiple implementation variants and tunable knobs for each operator,
including expensive but often overlooked operations such as expression
evaluation, string processing, and multi-key sorting, and it profiles
intermediate data during query execution with lightweight statistics to
select implementations, tune knobs, and compress data on the fly.
Our evaluation shows that this design pays off: Eiger outperforms the
state-of-the-art cuDF library by up to 1.8$\times$ on the complete TPC-H
benchmark and up to 6.1$\times$ on individual queries.
Beyond raw performance, our analysis characterizes how each variant and
configuration behaves across workloads and GPU architectures, providing a
foundation for cost models in future GPU query optimizers.
We hope that Eiger helps build an understanding of what an adaptive
GPU-based library for data analytics should look like in the era of
composable database systems~\cite{kabic25-maximus,velox-manifesto}.

\begin{acks}
This work was supported by a grant from the Swiss AI initiative and the Swiss National Supercomputing Centre (CSCS) under project ID sm94 and a donation from NVIDIA Corporation. 
\end{acks}

%\clearpage

\bibliographystyle{ACM-Reference-Format}
% \bibliography{sample}

\end{document}